\documentclass{article}

\usepackage{arxiv}

\usepackage[utf8]{inputenc} % allow utf-8 input
\usepackage[T1]{fontenc}    % use 8-bit T1 fonts
\usepackage{hyperref}       % hyperlinks
\usepackage{url}            % simple URL typesetting
\usepackage{booktabs}       % professional-quality tables
\usepackage{amsfonts}       % blackboard math symbols
\usepackage{amsmath}
\usepackage{nicefrac}       % compact symbols for 1/2, etc.
\usepackage{microtype}      % microtypography
\usepackage{cleveref}       % smart cross-referencing
\usepackage{graphicx}
\usepackage{doi}
\usepackage{import} % TODO remove?
\usepackage{cite}
\usepackage{orcidlink}
\usepackage[squaren,binary]{SIunits}
\usepackage[ruled,vlined,linesnumbered]{algorithm2e}
\usepackage{subfig}
\usepackage{multirow}

\DeclareMathOperator*{\argmax}{arg\,max}

\title{Just a Second -- Scheduling Thousands of Time-Triggered Streams in Large-Scale Networks}

\author{\orcidlink{0000-0002-5146-0184}Heiko~Geppert\\%\thanks{Use footnote for providing further
	IPVS\\
	University of Stuttgart\\
	Germany\\
	\texttt{heiko.geppert@uni-stuttgart.de} \\
	\And
	\orcidlink{0000-0002-3470-7712}Frank~D\"urr\\
	IPVS\\
	University of Stuttgart\\
	Germany\\
	\texttt{frank.duerr@uni-stuttgart.de} \\
	\And
	\orcidlink{0000-0001-5045-4252}Sukanya~Bhowmik\\
	IPVS\\
	University of Stuttgart\\
	Germany\\
	\texttt{sukanya.bhowmik@uni-stuttgart.de} \\
	\And
	\orcidlink{0000-0001-8986-8241}Kurt~Rothermel\\
	IPVS\\
	University of Stuttgart\\
	Germany\\
	\texttt{kurt.rothermel@uni-stuttgart.de} \\
}

\renewcommand{\shorttitle}{Just a Second}

\hypersetup{ 
pdftitle={Just a Second - Scheduling Thousands of Time-Triggered Streams in Large-Scale Networks},
pdfsubject={cs.NI},
pdfauthor={Heiko Geppert et al.},
pdfkeywords={deterministic networks, stream scheduling, TTEthernet},
}

\begin{document}
\maketitle

\begin{abstract}
	Deterministic real-time communication with bounded delay is an essential requirement for many safety-critical cyber-physical systems, and has received much attention from major standardization bodies such as IEEE and IETF.
	In particular, Ethernet technology has been extended by time-triggered scheduling mechanisms in standards like TTEthernet and Time-Sensitive Networking.
	Although the scheduling mechanisms have become part of standards, the traffic planning algorithms to create time-triggered schedules are still an open and challenging research question due to the problem's high complexity.
	In particular, so-called plug-and-produce scenarios require the ability to extend schedules on the fly within seconds.
	The need for scalable scheduling and routing algorithms is further supported by large-scale distributed real-time systems like smart energy grids with tight communication requirements.
	In this paper, we tackle this challenge by proposing two novel algorithms called Hierarchical Heuristic Scheduling (H2S) and Cost-Efficient Lazy Forwarding Scheduling (CELF) to calculate time-triggered schedules for TTEthernet.
	H2S and CELF are highly efficient and scalable, calculating schedules for more than 45,000 streams on random networks with 1,000 bridges as well as a realistic energy grid network within sub-seconds to seconds. 
\end{abstract}

% keywords can be removed
\keywords{deterministic networks \and stream scheduling \and TTEthernet}

% TODO make it all self contained
% !TeX spellcheck = en_US
% !TeX root = ../h2s_main.tex

\section{Introduction}
\label{sec:introduction}

Reliable and deterministic communication with bounded latency is an essential requirement for distributed real-time systems, especially when a violation of the real-time properties can cause major harm.
Prominent examples of such real-time systems include applications from the Industrial Internet of Things (IIoT) such as motion control of robots, or monitoring and controlling smart grids.

The great efforts of major standardization bodies and vendors to integrate deterministic communication capabilities into standard networking technology showcase the increasing importance of deterministic communication.
Notable examples include the Time-Sensitive Networking (TSN) extensions defined by the TSN Task Group of the IEEE 802.1 Working Group, TTEthernet (TTE), Ethercat, and SERCOS III.
They are all real-time Ethernet technologies for local and metropolitan area networks (LAN/MAN), but also serve as a basis for networks beyond the LAN/MAN scope.
In particular, deterministic communication is also discussed by the IETF DetNet Working Group for routed networks and for wireless 5G systems under the term Ultra-Reliable and Low Latency Communication (URLLC) \cite{comSocBlog}. 

In this paper, we focus on time-triggered (TT) scheduling driven by timetables in Ethernet bridges to give streams access to outgoing links at precisely defined points in time according to a given, pre-calculated timetable.
Such TT scheduling mechanisms are often applied to deterministic communication with very high demands on latency, jitter, and loss.
Isochronous traffic, in particular, poses challenges on scheduling mechanisms.
It is defined in the IEC/IEEE 60802 \cite{IECSC65C/WG18} as periodic traffic with short cycle times from \unit{100}{\micro\second} to \unit{2}{\milli\second} with fixed small application data sizes fitting in a single Ethernet frame, zero loss tolerance, short deadlines, and zero interference with other streams.
In addition, it is transmitted by senders that are synchronized to the network time, i.e., sending times and network schedules executed by bridges are synchronized.
TTEthernet or the TSN Time-Aware Shaper (IEEE 802.1Qbv) are prominent examples of such time-triggered schedulers. 
Although TT scheduling has been subject to standardization, these standards typically only cover the scheduling mechanisms, which execute scheduling according to a given timetable.
The algorithms required to calculate them are not included.

For planning TT schedules, several approaches have been proposed in literature \cite{Abbasloo2018,Behera2022,Finzi2022,Pahlevan2018,Pahlevan2019,Vlk2022,Wisniewski2015}.
Early endeavors considered the global scheduling problem with a priori given static sets of streams \cite{Duerr2016}.
Later work also considered the extended joint routing and scheduling problem, where routes to isolate streams in the space domain have to be calculated in addition \cite{Schweissguth2017}.
Typical methods applied here are Integer Linear Programming (ILP), Constraint Programming (CP) or other black-box solvers \cite{Hellmanns2020,Nayak2016,Pop2016,Santos2019,Steiner2010}.

Huge problem sizes arise from smart factories, coordinating thousands of sensors \cite{KUKA2016}, or energy grids spanning continents.
In all examples, a failure can easily harm humans or lead to serious financial loss.
However, established offline computations using ILP/CP solvers become infeasible considering these modern problem sizes.
Instead, fast and scalable heuristics are employed to compute the scheduling tables within seconds.
As a consequence, these solutions can be used online in dynamic environments where incremental system changes such as continuously arriving streams require modifications in the scheduling tables on the fly \cite{Tang2019,Huang2021}.
A frequently applied strategy for incremental scheduling (and optionally also routing) is to treat already added streams as fixed parameters.
Thus, only the time-windows and routes for newly added streams have to be defined as variables and solved in the incremental problem \cite{Falk2022,Nayak2016,Raagaard2017}.
Since already scheduled streams are not re-scheduled, this is typically referred to as a defensive planning strategy.
On the one hand, this defensive strategy avoids the issue of jitter during the transition from an old plan to a new plan.
On the other hand, fixing streams early might turn out to be suboptimal in comparison to solutions where existing schedules and routes can be altered later.
Despite this disadvantage, we advocate for defensive planning as the natural strategy for isochronous traffic since it prevents reconfiguration jitter by design.

In this paper, we propose a novel defensive algorithm for the incremental planning of schedules and routes for isochronous traffic in TTEthernet with three notable features.
Firstly, we aim to maximize the aggregated throughput of streams admitted to the network, which has so far mostly been neglected.
This objective has previously only been considered in DetNet environments \cite{Krolikowski2021}.
In contrast, approaches that target Ethernet often maximize the number of streams admitted, which benefits streams with long periods and small frame sizes.
Maximizing the aggregated throughput ties the solution assessment directly to the amount of data being transmitted.
Moreover, it improves the utilization of precious network resources (bandwidth) \cite{Heilmann2019, Zhang2020a}.
Although high-data-rate Ethernet technologies exist with tens to hundreds of \giga\bit\per\second, these are expensive and consume significantly more energy.
Moreover, they are difficult to use in certain environments such as a shop floor, where ruggedized passively cooled devices are preferable over devices that must be cooled with fans which collect dust and require a lot of space \cite{Vitturi2015}. 
Therefore, \unit{1}{\giga\bit\per\second} and \unit{100}{\mega\bit\per\second} devices are still common in such environments.

Secondly, our planning algorithm is based on heuristics that are highly efficient and can calculate schedules in sub-seconds or seconds on very large networks.
This facilitates plug-and-produce scenarios in which users expect a similar experience as in traditional networks where devices should function almost immediately when connecting them \cite{Prinz2018}.
While the solution presented in \cite{Bujosa2022} can be computed in similar time, it's per network link based scheduling method confines the solution to acyclic networks, since it can run into deadlocks on more complex topologies.
We address these limitations by applying a per stream based scheduling method, using multiple computationally cheap heuristics to navigate the solution space.
First, we decide on the next stream to be placed in the network, based on a stream sorting heuristic, e.g., selecting the stream with the smallest period first.
Next, we decide the route to be used, based on a route sorting heuristic, e.g., taking the shortest route with left over capacity.
In addition, we incorporate a mechanism to support partial solutions.
The Idea is to avoid rejecting all new streams by rejecting only a subset of them, while admitting the remaining ones.

Thirdly, our algorithms support buffering of frames at network bridges. 
Compared to approaches that use no-wait scheduling with zero buffering, previous work on TSN scheduling has found that supporting buffering results in higher admission rates \cite{Santos2019,Vlk2022}.

In detail, we make the following contributions:

\begin{itemize}
	\item H2S: A fast heuristic to solve the incremental scheduling problem, capable of scheduling up to  45,000 streams with a throughput of \unit{494}{\giga\bit\per\second} in a network with 1,000 bridges within sub-seconds to seconds.
	\item CELF Scheduling: A modification of H2S, with the ability to admit more streams, reducing the selection bias of H2S, at the cost of a slightly higher runtime.
	\item An extensive simulation-based scheduling evaluation on different topologies, including a realistic network model of a large smart grid, and a comparison with state-of-the-art scheduling solutions.
\end{itemize}

The remainder of the paper is structured as follows.
We cover the related work in Sec.~\ref{sec:related-work}, followed by our system model manifesting our network assumptions in Sec.~\ref{sec:system-model}.
Afterwards, we discuss our problem statement in Sec.~\ref{sec:problem-statement}.
In Sec.~\ref{sec:algorithms} we present our scheduling heuristic as well as the modified version.
Then, we evaluate them in Sec.~\ref{sec:evaluation}, and finally conclude the paper in Sec.~\ref{sec:conclusion}.

% !TeX spellcheck = en_US
% !TeX root = ../h2s_main.tex

\section{Related Work}
\label{sec:related-work}

The related work in the area of scheduling time-triggered traffic for real-time Ethernet can be structured into two classes:
Offline scheduling approaches that schedule a static set of a priori given streams, and online scheduling approaches that incrementally adapt schedules to a dynamic set of streams modified at runtime.
Note that rate constrained and best effort traffic is not a focus, even though some publications try to optimize their TT traffic in order to have capacity for the other traffic types.

Many established solutions focus on offline scheduling assuming all communications are known beforehand.
Popular methods include Integer Linear programming (ILP), Satisfiability Modulo Theories (SMT), and Constraint Programming (CP) \cite{Steiner2010,Craciunas2016,Pop2016,Santos2019,SernaOliver2018,Falk2018,Vlk2021,Minaeva2021}.
Besides scheduling---which is already an NP-hard problem by itself---, joint routing and scheduling problems have been investigated \cite{Schweissguth2017,Schweissguth2020}.
However, although ILP/SMT/CP solvers have evolved into very powerful tools, finding exact solutions for NP-hard problems is practically intractable for large-scale scenarios \cite{Hellmanns2020,SernaOliver2018}.

Therefore, heuristics became an important research focus to increase scalability of offline scheduling.
Using meta-heuristics such as tabu-search, the time to find feasible solutions can be reduced from days to hours or minutes \cite{Duerr2016,Pop2016,TamasSelicean2012}.
The Heuristic List Scheduler (HLS) schedules and routes 100 streams for small networks even within seconds, also supporting multicast communication and inter-stream dependencies \cite{Pahlevan2019}.
Nonetheless, the achieved runtimes are still far from the requirements of large-scale plug-and-produce scenarios where new schedules need to be available within seconds even for large networks. 

% Ragaard2017
Raagaard et al. presented a fast scheduling heuristic for TSN, capable of scheduling batches of several hundred streams within a few seconds \cite{Raagaard2017,Raagaard2017a}.
Similar to our approach, streams are added greedily in each iteration.
A defensive planning strategy is applied in a first step that adds new streams without re-scheduling already scheduled streams.
Only if the defensive scheduling fails, the approach falls back to calculating a completely new schedule, possibly also re-scheduling old streams.
In contrast to our approach, Raagaard et al. minimize the number of queues used for time-triggered traffic to maximize the number of queues for non-time-triggered traffic (Credit-Based Shaper or best-effort traffic).
Moreover, it does not support different candidate routes per stream or scheduling subsets of streams if insufficient network resources are available for scheduling all requested streams. 

%Vlk
Vlk et al. propose a heuristic similar to that of Raagaard et al., which implements so-called back-jumps to avoid local minima \cite{Vlk2022}.
This algorithm can handle large network topologies and schedule thousands of streams within minutes in most cases.
However, due to the focus on very large numbers of streams and runtimes in the minutes range, this approach is better classified as a scalable offline scheduling approach rather than incremental online scheduling.

% IRAS -> ~1s per stream
The Incremental Routing and Scheduling (IRAS) \cite{Huang2021} is a clear online scheduling algorithm that adds single streams to the schedule one by one.
For a set of candidate routes, IRAS first checks if the new stream is trivially schedulable, and invokes an ILP if not.
The average runtime to add a single stream is below a second for the small network topologies that the authors used in their evaluation.
However, due to falling back to an ILP solver rather than a heuristic solution, adding single streams can take much longer in the worst cases.
Further, adding batches of streams and buffering is not supported.

% GFH
A very different method was proposed by Falk et al. in \cite{Falk2020,Falk2022}.
The routing and scheduling problem is mapped to an independent colorful set problem.
The scheduling and routing options are encoded as colored vertices in a conflict graph where the vertex color represents the stream and edges are placed whenever two scheduling options are not compatible.
The so-called Greedy Flow Heap heuristic (GFH) finds an independent colorful set, which then provides a zero-queuing schedule.
While GFH has small runtimes even for larger problem instances, the construction of the conflict graph can take orders of magnitude longer and requires a significant amount of memory, making it less attractive for fast online scheduling.

% FlexCurve
G\"artner et al. proposed a scheduling approach for dynamic routing and scheduling in TSN recently.
It explicitly strives for extensibility by introducing the so-called flexcurve concept to model the flexibility of schedules with respect to adding new streams in the future \cite{Gaertner2021,Gaertner2022}.
However, this work is focuses on the notion of flexibility rather than efficient scheduling heuristics.

% Hermes
In 2022, Bujosa et al. presented Hermes \cite{Bujosa2022}, a very fast stream scheduling heuristic that uses a per link scheduling approach instead of the more common per stream method.
They provide two different modes: zero and relaxed reception jitter.
Evaluations on small networks showed a runtime considerably below a second.
However, the Hermes algorithm does not support routing and can fail due to deadlocks in the stream routes, which effectively limits it to acyclic networks.

To summarize, many endeavors tackled different aspects of the scheduling problem, like buffering, scaling, routing, etc.
However, we have yet to see a fast and scalable algorithm, which allows for frame buffering, supports multiple candidate paths, and works independent of the network topology.
% !TeX spellcheck = en_US
% !TeX root = ../h2s_main.tex

\section{System Model}
\label{sec:system-model}

% network graph
We assume a switched network with (only) full-duplex links modeled as a graph $\mathcal{G}$.
Network elements (end stations and bridges) are represented as vertices ($\mathcal{V}$).
Directed edges $\mathcal{E}$ denote the links between network element ports.
Each egress port has an associated frame buffer for outgoing frames.
We consider a standard store-and-forward model, where bridges can forward a frame only after they have received it completely.
However, our approach could be easily adapted to cut-through switching, where frames can already be forwarded after having received the first bytes of the Ethernet header.
Further, we expect bridges to be capable of stream based scheduling, allowing for frames to overtake each other at bridges.
% bridges' storage size
Although, in practice bridges only have limited storage, modern TTE bridges can have \unit{512}{\kilo\byte} of frame memory\footnote{cf. TTE Switch Space 3U cPCI}.
This is sufficient for several hundred maximum transmission unit (MTU) size frames.
In our evaluated scenarios, the maximum number of buffered frames required at a single egress port was always small enough to fit into the memory.
Hence, all network requirements are met by TTEthernet.
When implementing a limit of 8 or fewer frames to be buffered at a single egress port, any solution could also be applied to IEEE 802.1Qbv networks and still ensure frame isolation.
However, this would significantly reduce the solution space, limit the scheduling success and prevent rate constrained or best effort traffic.
Thus, we focus on TTEthernet networks in this paper.

\subsection{Stream Definition}
% stream set
The time-triggered traffic is modeled by the stream set $\mathcal{S}$.
Every stream has a source node, a single destination node, frame size, period, and deadline which is equal to the period in our case.
Both source and destination are end stations in $\mathcal{V}$.
Periods of individual streams can be different, yet the periods should be chosen so that the hyper period is within reasonable limits, e.g., a few \milli\second . % this space is needed. Otherwise the dot is not displayed
Moreover, we expect the start of periods to be in-phase, which is the critical instant for the system.
Systems that do not comply naturally could be modified to delay the release of frames and use smaller periods to counteract the initial delay.
The deadlines are considered to be equal to the period, so that a message, e.g., sensor reading, becomes invalid as soon as there is a newer one.
Note that longer deadlines could interfere with the next hyper period's traffic, and therefore, best be handled as if the deadline was equal to the period.
Meanwhile, shorter deadlines would inherently waste a lot of bandwidth when assuming in-phase releases.

\subsection{Time Synchronization and Configuration Jitter}
% clocks and jitter
We assume that the clocks of all bridges and end stations are sufficiently synchronized to enable scheduling with \unit{1}{\micro\second} macro ticks, comparable to \cite{Craciunas2016}.
Note that although this can eliminate release time jitter, frames still might arrive at the destination with jitter from intermediate buffering according to the schedule.
This is because the queuing delay can be different for successive frames of the same stream.
We do not restrict this type of jitter because we have variable, but bounded, queuing delays.
Further, the recipients could add a small buffering to prevent this type of jitter, since the clocks of all devices are synchronized.
Instead, we focus on deterministically bounded end-to-end delays. 
Thus, the stream specification does not include a jitter bound, only a deadline.
Details on frame sizes and periods will be provided in Section~\ref{sec:evaluation}.

\subsection{Stream Isolation in Time and Space}
To isolate streams in time, each network element implements a time-triggered scheduling mechanism.
Therefore, they have scheduling tables defining when a buffered frame of a given stream is forwarded over an outgoing port, i.e., an individual schedule is defined for each outgoing port.
The essential requirement for a valid schedule is that the forwarding time windows assigned to each frame of a stream must not overlap since only one frame can be transmitted at a time over an outgoing port.

% isolation in space
Besides this isolation in the time domain, we also consider isolation in space through explicitly defined routes per stream.
To this end, bridges implement programmable forwarding tables to specify individual routes defined as a sequence of outgoing ports for each stream.
Both schedules and routes are defined by a central entity that has a global view onto the network topology and the set of streams, like a centralized network controller (CNC) \cite{IEEE802.1Qcc-2018}.
To admit a new stream $s$, the central entity computes \emph{candidate routes} and reserves \emph{time windows} (\emph{slots}) for all frames of $s$ on all egress ports along a selected candidate route, ensuring a timely arrival at the destination.
The length of the time window is defined by the frame size and the data rate of the port. 

\subsection{Dynamic Stream Updates}
% Dynamic scenario
The set of streams to be hosted in the network can be modified at runtime by adding batches of streams or removing batches of already scheduled streams. 
A batch update request consists of newly added streams $\mathcal{R}^{\text{add}}$ and removed streams $\mathcal{R}^{\text{del}}$, where the batch size can be arbitrarily large.
One of the batches can also be empty, e.g., when only new streams are added or only existing streams are removed.
The central entity collects stream requests to form batches as appropriate by the use case, which may be every x seconds, as soon as x requests are collected, or as soon as the previous batch is scheduled.
We assume larger batches, i.e., a new application is rolled out on a production line resulting in many stream updates in a single batch.
To ensure that the guarantees for already admitted streams are not violated, new streams can be rejected ($\mathcal{R}^{\text{rej}}$) by the scheduler, i.e., the set of actually added streams $\mathcal{R}^{\text{add}} \setminus \mathcal{R}^{\text{rej}}$ can be smaller than the set of requested streams to be added. 
In Definition~\ref{eq:stream_set_at_time_i}, $\mathcal{S}_i$ denotes the stream set successfully scheduled after the $i$-th batch update:

\begin{equation}
	\label{eq:stream_set_at_time_i}
	\mathcal{S}_i = \big (\mathcal{S}_{i-1} \setminus \mathcal{R}^{\text{del}}_i \big )\cup \mathcal{R}^{\text{add}}_i \setminus \mathcal{R}^{\text{rej}}_i
\end{equation}

\subsection{Update Protocol and Failure Model}
% Update protocol
Whenever a new configuration is computed, the bridges need to be updated.
Our stream model allows for a straightforward update scheme where the central entity distributes updated scheduling tables.
All bridges switch to the new tables at the same hyper period transition.
Since all frames need to be received before their deadline, and we assume the deadline equals the period, there is no time-triggered traffic left in the network during this transition.

Removing streams is quite simple due to the previously discussed update model.
However, removing streams leads to fragmentation in the schedule, which in turn degrades the overall network utilization over time.
Defragmentation strategies, i.e. offensive planning strategies, can counteract this issue at the cost of introducing reconfiguration jitter by rearranging previously admitted streams.

% Failure model
For now, we assume a simple failure model without automatic network recovery in case of link or node failures.
This means that, in case of an error, some destinations could become unreachable, and the remaining links might not have sufficient capacity to reroute all affected streams.
Handling these problems is beyond the scope of this paper.

% !TeX spellcheck = en_US
% !TeX root = ../h2s_main.tex

\section{Problem Statement}
\label{sec:problem-statement}

The problem to be solved is to incrementally calculate traffic plans consisting of per-stream schedules and routes that optimize the aggregated network throughput of the admitted streams.

More formally, we define an incremental planning function $\mathrm{Plan}(\mathcal{G}, \mathcal{S}_i, \mathcal{R}_{i+1}) \mapsto \mathcal{S}_{i+1}$, which takes as input the network $\mathcal{G}$, the already admitted streams $\mathcal{S}_i$, including their schedules and routes, and a request batch $\mathcal{R}_{i+1}$. 
The request batch holds the streams to be added ($\mathcal{R}^{\text{add}}_i$) as well as the removed ones ($\mathcal{R}^{\text{del}}_i$).
We define a combination operator $\diamond$ to apply the request batches as shown in Definition~\ref{eq:stream_set_at_time_i}.
Thus, $\mathcal{R}_0 \diamond \mathcal{R}_1$ represents all flows to be added so far without the streams that left the system.

In contrast to related approaches, we propose an optimization criterion that maximizes the aggregated throughput of admitted streams, rather than maximizing the number of admitted streams.
This approach has the potential to remove the bias towards smaller streams.
Formally, let $\mathrm{thr}(s)$ be the throughput of a stream $s$ as shown in Definition~\ref{eq:throughput}.

\begin{equation}
  \label{eq:throughput}
  \mathrm{thr}(s) = \dfrac{\mathrm{framesize}(s)}{\mathrm{period}(s)}
\end{equation}

We then strive to find a feasible subset of streams $S \subseteq \mathcal{R}_1 \diamond \ldots \diamond \mathcal{R}_n$ that maximizes the aggregated throughput and fulfills the scheduling and routing constraints, i.e., no overlapping forwarding time windows, and routes connecting sources to destinations.
This is formalized in Definition~\ref{eq:optimal_solution}:

\begin{equation}
	\label{eq:optimal_solution}
	\argmax_{S \subseteq \mathcal{R}_1 \diamond \ldots \diamond \mathcal{R}_n} \sum_{s \in S}\mathrm{thr}(s) \qquad \textrm{s.t. $S$ is feasible}
\end{equation}

An optimal solution needs to yield maximum aggregated throughput after the very last time step.
This corresponds to an offline planner to which all requests are known a priori.
We have chosen Definition~\ref{eq:optimal_solution} as optimization criterion because we believe that incremental planning algorithms should incorporate mechanisms that account for the existence of future requests.
Practically, the scheduler does not know the number of time steps, and in dynamic systems we have to assume infinite time steps.
Therefore, the scheduler has to perform well in every step, and it might schedule streams in a way that preserves the flexibility for admitting future requests.

% Terminology Table
\begin{table}
	\centering	
	\small
	\renewcommand{\arraystretch}{1.1}
	\begin{tabular}{c|c}
		$\mathcal{G}$& Network Graph \\
		\hline
		$\mathcal{V}$& End stations and bridges in $\mathcal{G}$ \\
		\hline
		$\mathcal{E}$& Network links in $\mathcal{G}$ \\
		\hline
		$\text{thr}(s)$ & throughput of stream $s$ (Definition~\ref{eq:throughput})\\
		\hline
		$\mathcal{S}_i$& Streams scheduled at time $i$ (Definition~\ref{eq:stream_set_at_time_i}) \\
		\hline
		$\mathcal{R}_{i}$ & Request set at time $i$. Consists of $\mathcal{R}_i^{\text{add}}$ and $\mathcal{R}_i^{\text{del}}$ \\
		\hline
		$\mathcal{R}_i^{\text{add}}$ & Streams to be added at time $i$ \\
		\hline
		$\mathcal{R}_i^{\text{del}}$ & Streams to be removed at time $i$ \\
		\hline 
		$R_i^{\text{rej}}$ & Streams rejected at time $i$. Subset of $\mathcal{R}_i^{\text{add}}$.\\
		\hline
		SSF & Stream Sorting Function (Definition~\ref{eq:ssf})\\
		\hline
		RSF & Route Sorting Function \\
		\hline
		CRF & Celf Rating Function (Definition~\ref{eq:crf}) \\
		\hline
		$\alpha$ & large constant, $\alpha > \max \text{framesize} \land \alpha > \max \text{ID}$ \\
	\end{tabular}
	\caption{Notation Overview}
\end{table}

% !TeX spellcheck = en_US
% !TeX root = ../h2s_main.tex

\section{Stream Scheduling}
\label{sec:algorithms}

% outlook on the subsections
In the following, we present our Hierarchical Heuristic Scheduling (H2S) algorithm to solve \emph{Plan} using two heuristics to navigate the solution space of stream and candidate route combinations.
This includes a discussion of H2S's algorithmic complexity. 
Finally, we show an adaptation of H2S based on the cost-efficient lazy-forwarding principle (CELF).
The CELF approach reduces some issues arising when using hierarchical heuristics by incorporating multiple heuristics into a single one.

% overview of the algorithm
The overall workflow starts with the central entity that collects add stream requests forming a batch.
For each new stream, a set of candidate routes is computed (cf. Appendix~\ref{sec:routing}).
Afterwards, \emph{Plan} is executed on the batch using the strategy that we discuss in the following.
The resulting schedule is then rolled out in the network.

\subsection{H2S: Hierarchical Heuristic Scheduling Algorithm} 
\label{sec:h2s}

The Hierarchical Heuristic Scheduling algorithm solves the \emph{Plan}($\mathcal{G}, \mathcal{S}_i, \mathcal{R}_{i+1}$) function discussed in Section~\ref{sec:problem-statement}.
H2S takes a given network graph $\mathcal{G}$, previously scheduled streams $\mathcal{S}_i$ and the next request batch $\mathcal{R}_{i+1}$.
H2S solves \emph{Plan} greedily in a fast one-pass assignment by consecutively adding streams to the network based on a \emph{Stream Sorting Function} (SSF) and using a candidate route according to the \emph{Route Sorting Function} (RSF).
Thereby, short periods and short distances are preferred properties for streams and routes because they lead to a lower reduction of the solution space and overall network utilization.
Every intermediate step yields a valid, yet not necessarily optimal, solution.

\begin{algorithm}
	\KwIn{$\mathcal{G}, \mathcal{S}_i, \mathcal{R}_{i+1}$}
	\KwResult{$\mathcal{S}_{i+1}$}
	$\mathcal{S}_{i+1} \gets \mathcal{S}_i$\;
	$heap_{stream} \gets \textit{heapify}(\mathcal{R}_{i+1},\text{SSF})$\;
	\While{$heap_{stream} \neq \emptyset$}{
		$s \gets \text{heap}_{\text{stream}}\text{.pop()}$\;
		$heap_{route} \gets \textit{heapify}(P^s, \text{RSF})$\;
		\While{$heap_{route} \neq \emptyset$}{
			$p \gets heap_{route}\text{.pop()}$\;
			\If{place($\mathcal{G}, s, p$)}{ \label{alg:h2s:place}
				$\mathcal{S}_{i+1} \gets \mathcal{S}_{i+1} \cup s$\;
				\textbf{break}\tcp*{goto next stream}
			}
		}
	}
	\KwRet{$\mathcal{S}_{i+1}$}\;
	\caption{H2S Algorithm}
	\label{alg:h2s}
\end{algorithm}

The pseudocode for H2S is given in Algorithm~\ref{alg:h2s}.
First, the result set $\mathcal{S}_{i+1}$ is initialized with $\mathcal{S}_i$.
Next, the streams from $\mathcal{R}_{i+1}$ are stored in a heap ($\textit{heap}_{\textit{stream}}$), ordered by the SSF, which prioritizes short periods, aiming for a high total aggregated throughput.
The streams in $\textit{heap}_{\textit{stream}}$ are iterated and in every iteration, the top stream $s$ is taken.
For each $s$, H2S orders the candidate routes $\mathcal{P}^s$ into $\textit{heap}_{\textit{route}}$ based on the RSF.
We try to schedule $s$ using the first (shortest) candidate route, queuing frames whenever necessary on the route.
If the placement (cf. line~\ref{alg:h2s:place}) succeeds, the network resources are reserved and H2S continues with the next stream from $\textit{heap}_{\textit{stream}}$.
Otherwise, it tries the next best candidate route from $\textit{heap}_{\textit{route}}$.
In case $s$ cannot be scheduled using any route in $\mathcal{P}^s$, $s$ is rejected and H2S continues with the next stream.

% place method
Besides checking for feasibility of a stream-route combination and reserving the network resources along the stream's route, the \emph{place} method also tries to preserve capacity for unknown future streams. 

The issue with plain ASAP placements, as in \cite{Pahlevan2019,Raagaard2017}, is that the start of the hyper period is saturated quickly.
Thus, there is no capacity left for the first frame of new streams if they have short periods (and therefore deadlines).
As a result, they have to be rejected, because the first frame would miss its deadline.
Our solution is to space out the placements, so that we have capacities left throughout the hyper period for as long as possible.

\begin{figure}
	\centering
	\includegraphics[width=0.6\linewidth]{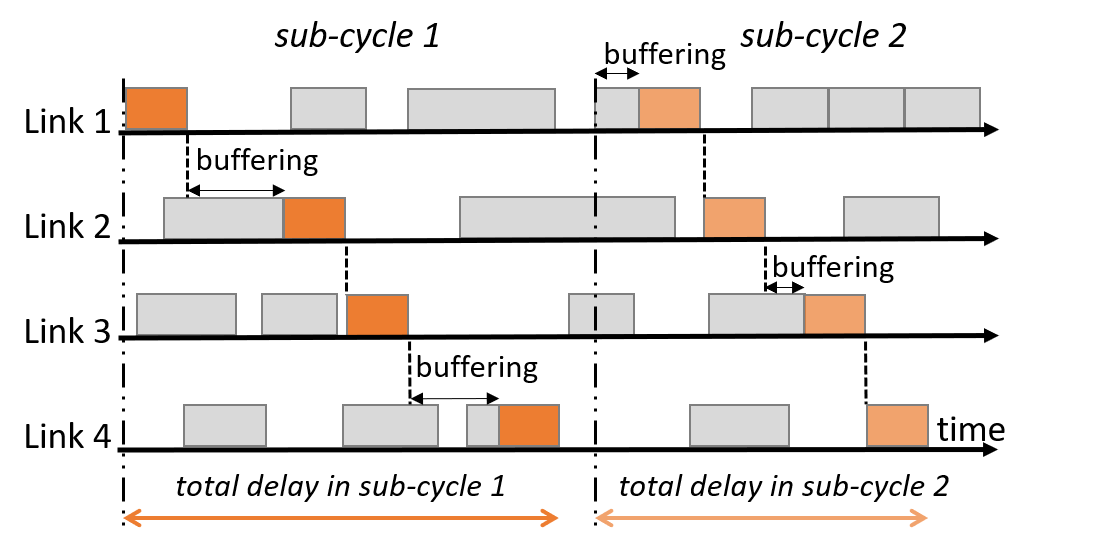}
	\caption{Stream placement example. The orange transmissions are possible forwardings for the new stream, the gray transmissions are reservations from previously placed streams.} 
	\label{fig:flow-placement}
\end{figure}

Checking every possible initial offset would take a lot of time.
Therefore, we only take some offsets which we call sub-cycles, based on the periods in the system.
If we knew about future stream periods, we could also include those in the decision.
We define a sub-cycle as the greatest common divisor of all known periods. 
When using harmonic or almost harmonic periods, this is the smallest period in the system.
To place a stream, we check the accumulated total delay (start of the sub-cycle until the frame is delivered) for every sub-cycle and then take the one with the smallest delay.
An example is shown in Figure~\ref{fig:flow-placement}, consisting of two sub-cycles.
We try to place a stream traversing the links 1 to 4.
The gray boxes represent capacities reserved by other streams.
The placements within the sub-cycles are shown in orange.
Whenever a frame can not be forwarded immediately it is buffered.
This can also happen at the start of a sub-cycle (cf. sub-cycle 2).
We decide to schedule the stream with an offset starting at sub-cycle 2, because the total delay is lower there.
Finally, the network resources are reserved for all links.

Note that, in reality, the decision becomes slightly more complicated, since we have to check the delay for all frames within one hyper period.
Here, we use the highest total delay of a frame to decide the sub-cycle.

Our load balancing adaptation yielded small improvements at a fairly low cost in preliminary evaluations.
However, investigating additional placement strategies might be beneficial.
We defer this to future work, since it is not the focus of this paper.

In the following, we discuss the heuristics used to sort the streams and routes in more detail.
Note that the heuristics are exchangeable if other optimization goals are to be achieved.
The presented heuristic functions proved to yield good results when optimizing the aggregated throughput.

\paragraph*{Stream Sorting Function}
\label{sec:H2S:heuristics:stream_selection}

Previously, we discussed the impact of the periods on the available capacity. 
To further circumvent the issue of unavailable capacities during early periods, we try to schedule low period streams first.
Otherwise, the egress ports are completely utilized in the early stage of the hyper period by streams with long periods and adding low period streams fails because the first frame misses its deadline.
This is also in line with the basic idea of rate monotonic scheduling, or, since we assume period equals deadline, the scheduling decision from \cite{Raagaard2017}.
Hence, the streams with small periods are scheduled first and the ones with longer periods ``fill up'' the gaps later.
As a first tie-breaker, we use the frame size, benefiting streams with large frame size since ``small'' streams might fit in between others later while ``larger'' streams might not.
The final tie-breaker is the ID, e.g., given by FIFO ordering, making it a deterministic heuristic.
Technically, we assign every stream a real number and use them to sort the streams in ascending order.
This ``score'' is computed as follows:

\begin{equation}
	\label{eq:ssf}
	SSF(x) \rightarrow \mathbb{R}: \alpha * \text{period}(x) + \frac{\alpha}{\text{frame size}(x)} + \frac{\text{id}(x)}{\alpha}
\end{equation}

Parameter $\alpha$ denotes a large constant to ensure that the first term, built from the period, is always larger than the second term, based on the frame size.
We used $\alpha=10,000$, but smaller values can be used as long as $\alpha$ is larger than the largest frame size and highest ID.
The second term in turn is more significant than the third term which is derived from the ID. 
In other words, we ensure that no matter how small the frame size is, a period which is a single micro tick longer will always dominate the score.

\paragraph*{Route Sorting Function}
\label{sec:H2S:heuristics:route_selection}

After selecting the next stream to assign, we need to decide on its candidate route.
All frames have to arrive within the deadline using the selected route.
Further, network congestion should be avoided, so future streams can still be scheduled.
Note that the route sorting heuristics are not routing algorithms, they only sort the candidate set of possible routes.
The routing algorithm to calculate candidate routes is given in Appendix~\ref{appendix:routing}.

Shortest paths are often used in the literature \cite{Falk2020,Falk2022,Pop2016}.
Using short routes results in smaller network utilization and deadlines might be met implicitly.
Therefore, ordering the candidate routes by their best case delay reduces the overall utilization which allows for additional streams in the future.
In our homogeneous networks where all links have the same propagation and transmission delay, and all bridges have the same processing delay, we can simply use the hop distance.
Further, the shortest route is a beneficial metric since it can be precomputed and cached for later use.

\paragraph*{Complexity Analysis}

H2S has multiple building blocks to be considered in the complexity analysis: the creation of the stream heap and route heaps, the sorting functions for the stream and route selection, and finally the actual insertion of a stream into the schedule.
We can create the heap in $\mathcal{O}(n)$ and have a removal complexity of $\mathcal{O}(\log n)$.
The complexity of both rating heuristics for a single stream/route evaluation is $\mathcal{O}(1)$ since it is a simple lookup of the stream properties or route length.
Creating and iterating the stream heap is an $\mathcal{O}(\mathcal{R} \log \mathcal{R})$ operation.
The route selection heap is built and iterated in $\mathcal{O}(c \log c)$, which is done $\mathcal{R}$ times.
Thereby, $\mathcal{R}$ denotes the number of streams added in a batch $\mathcal{R}_{i+1}$ and $c$ is the number of candidate routes per stream.
Placing a selected stream (independent of the success) takes at most $\mathcal{O}(\hat{p} \cdot h)$ time, where $\hat{p}$ is the hop distance of the longest candidate route and $h$ the length of the hyper period.
The number of insertion attempts is upper bounded by $c*\mathcal{R}$ but in most cases quite close to $\mathcal{R}$.
Further, we do not need to perform complex computations for every macro tick in the hyper period but check for sufficiently sized free slots, relating to a much smaller factor than $h$.
Combining all these complexities results in $\mathcal{O}(\mathcal{R}(\log \mathcal{R} + c \log c + c \cdot \hat{p} \cdot h))$ time in the worst case, which is less than the routing in large networks for $\mathcal{R}$ streams with Dijkstra's algorithm.

\subsection{CELF-Scheduling Adaptation}
\label{sec:celf-scheduling}

The hierarchical approach of the H2S algorithm has the benefit of a very fast assignment, because it uses multiple simple choices.
However, this can lead to obviously bad assignments, instead of rejecting a single stream in order to admit many streams later.
To be more precise, H2S might be forced to schedule the current stream via an undesirable route, e.g. taking the long route in a ring network, because the SSF decided to place the current stream.
This becomes especially apparent in scenarios where the network is over-saturated and the stream candidate routes differ a lot in quality.
The Cost-efficient Lazy-forwarding scheduling (CELF) is an adaptation of H2S which reduces this problem at the cost of higher runtimes.
To avoid undesirable placements, CELF does not pick a specific stream to schedule next whatever the cost, but tries to schedule the best stream-route combination using a single hierarchical scoring heuristic instead of multiple heuristics hierarchically.

The algorithm adopts the cost-efficient lazy-forwarding technique \cite{Leskovec2007}, which is well known in the area of influence maximization \cite{Kim2013,Tang2017,Geppert2021}.
The idea is to use the monotonicity and submodularity of a scoring function to reduce the number of recomputations needed due to decisions taken during the algorithm.
To be more precise, every stream we schedule might impact the ``score'' our heuristic assigns to the remaining stream-route combinations.
If the score only decreases over time (we strive for high scores), we can use previously computed scores as upper bounds and recompute the score only if required.
In practice, we can compute the score for every stream-route combination once and sort the whole list. 
After scheduling and removing the best option, we can recompute the new best score and if the new result is still better than the old score of the new second-best option, we know for sure that there is no other option with a higher score.
This is because the scores only decrease, and the old scores can still be used as upper bounds.
In case the new result was worse than the old score of the second-best option, the newly computed score is reinserted in the sorted list, and we repeat the procedure.

\begin{algorithm}
	\KwIn{$\mathcal{G}, \mathcal{S}_i, \mathcal{R}_{i+1}$}
	\KwResult{$\mathcal{S}_{i+1}$}
	$\mathcal{S}_{i+1} \gets \mathcal{S}_i$\;
	CRF $\gets$ Celf rating function\;
	$heap \gets \textit{heapify}(\langle s',p'\rangle \forall p' \in P^{s'} \forall s' \in \mathcal{R}_{i+1},\text{CRF})$\;
	\While{$heap \neq \emptyset$}{
		$\langle s,p\rangle \gets \text{heap}_{\text{stream}}\text{.pop()}$\;
		\If{$s \in \mathcal{S}_{i+1}$}{
			\tcp{stream is already placed}
			\textbf{continue}\;
		}
		\If{place($\mathcal{G}, s, p$)}{
			\tcp{successful stream placement}
			$\mathcal{S}_{i+1} \gets \mathcal{S}_{i+1} \cup s$\;
			
		}
	}
	\KwRet{$\mathcal{S}_{i+1}$}\;
	\caption{Celf-Scheduling}
	\label{alg:celf}
\end{algorithm}

Our CELF adaptation, shown in Algorithm~\ref{alg:celf}, provides a different strategy to solve \emph{Plan} than H2S.
Instead of a stream sorting function and a route sorting function it has one rating function, assigning a score to stream-candidate route combinations used to order the combinations into a max-heap.
After the initial ordering, the stream-candidate route options are iterated and the algorithm removes the first candidate and tries to schedule the stream via the candidate route using the same \emph{place} function as H2S.
If it succeeds, all other candidate routes of the same stream can be ignored or removed.
When the next stream-candidate route combination is taken from the heap, its score is recomputed.
In case the new score is the best compared to the maybe outdated score of the next pair, then the algorithm can continue with the stream-candidate route pair as described before.
Otherwise, the current pair reenters the heap with its updated score and the process is repeated.
The heap is iterated until all nodes are either placed, removed because placement is not possible, or ignored since the nodes' streams are already scheduled.

\paragraph*{CELF Rating Function}

From preliminary evaluations of H2S and \cite{Raagaard2017} we know that scheduling streams with small periods early is important for the overall scheduling success.
However, using always the shortest route can be suboptimal \cite{Nayak2018} and considering the utilization to prevent bottlenecks can be efficient \cite{Syed2021}.
While using short routes typically reduces the overall load, the stream rating function can become an issue if no short route are unavailable due to congestion.
H2S, for example, would try to schedule a stream via the long route in a ring if the short one is unavailable, instead of scheduling multiple other streams on these links, which in turn can reduce the final aggregated throughput.
The CELF rating function orders all stream-candidate route pairs based on 1)~their period (small periods earlier) 2)~their route rating in terms of the route's utilization, and 3)~their IDs as tiebreaker:
\begin{equation}
	\label{eq:crf}
	\textit{CRF}(s,p) = \frac{\alpha}{\text{period}(s)} + \frac{1}{1 + \sum_{l \in p}\text{utilization}(l)} + \frac{\text{id}(s)}{\alpha}
\end{equation}
Again, $\alpha$ is a large constant ensuring that earlier terms dominate the later terms (we used 10,000 cf. Definition~\ref{sec:H2S:heuristics:stream_selection}).
To ensure that $\alpha$ is the dominating factor it needs to be larger than twice the hyper period.
Further, $\alpha \geq max~id \land \alpha \geq max~\left |route \right |^2$ needs to hold, where $max~id$ is the highest id and $max \left |route\right |$ is the hop distance of the longest candidate path.
The route rating is the sum of the utilization of every link $l$ in the candidate route $p$.
This way, unused routes are preferred, but short routes are still selected often since they have fewer utilization values to sum.

\paragraph*{Complexity Analysis}
The basic building blocks of the CELF-Scheduling are the rating function computation, the heap structure management, and the stream insertion into the schedule.
Computing the CRF for a single stream-candidate route pair takes $\mathcal{O}(\hat{p}\cdot h)$ time when we compute the network links' utilization on demand.
Again, $\hat{p}$ denotes the longest candidate route and $h$ the length of the hyper period.
The candidate rating needs to be computed $\mathcal{R}^2\cdot c^2$ times in the worst case and the heap management introduces another $\mathcal{R}\cdot c \log (\mathcal{R} \cdot c)$ overhead when the scores are cached.
As before, the stream placement is an $\mathcal{O}(\hat{p}\cdot h)$ operation.
This results in an accumulated time complexity of $\mathcal{O}(\mathcal{R}^2\cdot c^2\cdot\hat{p} \cdot h + \mathcal{R} \cdot c \log (\mathcal{R} \cdot c))$.

\subsection{Offensive Planning}

So far, we covered both H2S and CELF as defensive planning strategies.
This way, they are universally applicable and do not induce any reconfiguration jitter, because past scheduling decision stay fixed.
However, in dynamic systems where existing streams leave the system and new ones enter, defensive planning strategies lead to fragmentation, which in turn leads to resource waste.
Hence, defragmentation algorithms, so-called offensive planning strategies are needed.

H2S and CELF can be executed as offensive strategies.
The overall approach is thereby comparable to the GFH scheduling \cite{Falk2022}.
The scheduler first runs H2S (or CELF) in a defensive mode.
As long as all streams are admitted there is no need for offensive planning.
If a stream is not admitted and would be rejected by the defensive planning strategy, we run it in offensive mode.
Thus, we run the strategy again assuming an empty network, to be able to place the previous streams differently.
To uphold the quality of service guaranteed to the previous streams, we first run H2S (or CELF) with the intention to place all old streams.
As soon as this is done, it is invoked again trying to place the new streams.
Finally, the results of the defensive and offensive planning are compared and the better one is used as the next traffic plan.
Note that it can happen that the offensive strategy is unable to readmit all old streams.
In this case, the offensive solution is discarded.

% !TeX spellcheck = en_US
% !TeX root = ../h2s_main.tex
\section{Evaluation}
\label{sec:evaluation}

In this section, we present the evaluation of our scheduling heuristics.
Initially, we provide an overview of the evaluation setup.
We then simulate our approaches and several other scheduling strategies on different networks and communication scenarios, including a smart grid application to compare their performance.

%\subsection{Evaluation Setup}
%\label{sec:eval:environment}

% assumptions
In our evaluations, we assume store-and-forward bridges with a processing delay of \unit{4}{\micro\second} as in \cite{Schweissguth2020}.
Further, we expect all links in the networks to have a bandwidth of \unit{1}{\giga\bit\per\second} and propagation delay of \unit{1}{\micro\second} \cite{Falk2022}.
Frame sizes are randomly selected from $\unit{125}{\byte} \times \{1, 2, 4, 6, 8, 12\}$ if not stated otherwise.
Hence, each message fits into a single Ethernet frame.
The frame sizes are derived from the macro ticks, i.e., it takes \unit{1}{\micro\second} to transmit \unit{125}{\byte} on a \unit{1}{\giga\bit\per\second} network link, and \unit{12}{\micro\second} for \unit{1500}{\byte}.
With smaller frames, we would waste some bandwidth, leading to a lower aggregated network throughput.
Alternatively, one could increase the precision by using smaller macro ticks, which in turn would increase the overall computation time linearly, and is limited by the clock synchronization.

We consider the deadline of a stream to be equal to its period which is randomly picked from (\unit{250, 500, 1000, 2000}{\micro\second}).
The number of streams depends on the network size, but in most cases we investigate highly utilized networks.
The stream sources and destinations were chosen uniformly at random.
The scheduling granularity (macro tick) is \unit{1}{\micro\second}.
Our scheduling strategies were tested on different network topologies with a broad range of distinct properties such as diameters, numbers of possible candidate routes, etc.
We used multiple network instances when evaluating irregular topologies to avoid bias based on specific network instances.

% benchmarks
In the following, we provide a brief overview of the benchmark algorithms:
\begin{itemize}
	\item \textbf{FirstFit (FF):} simple first fit assignment using only the shortest route, a batch size of 1, and ASAP placement.
	Corresponds to H2S without any sorting logic.
	
	\item  \textbf{Earliest Deadline First (EDF):} offensive planning strategy running multiple EDF simulations with different subsets of $\mathcal{R}^{\text{add}}_i$ to find a feasible solution. 
	EDF has more degrees of freedom than most other algorithms, since it performs offensive planning, i.e. it is able to reschedule admitted streams (cf. Appendix~\ref{appendix:edf}).
	
	\item \textbf{Greedy Flow Heap Heuristic (GFH):} conflict graph-based, zero-queuing TSN scheduling heuristic \cite{Falk2022}, with use cases and evaluation scenarios comparable to ours.
	GFH is also capable of offensive planning but uses defensive planning as fall-back.
	We adapted GFH to our system model with strict deadlines. 
	We limited the configurations per stream, thus limiting the conflict graph size, and display the parameter as part of the name.

	\item \textbf{Hermes:} recent all-or-nothing stream scheduling heuristic with queuing and very fast runtimes \cite{Bujosa2022}. 
	We did not limit the number of queues to have a fair comparison.
	
	\item \textbf{Constraint Programming (CP):} Constraint Programming formulation with and without queuing (CP \& CP-0q) using the IBM ILOG CP solver (cf. Appendix~\ref{appendix:cp}). 
	The CP provides exact solutions when the runtime is unbounded.
	We set a limit of \unit{2}{\hour}. 
\end{itemize}

% environment
We implemented our scheduling algorithms in C++20 and performed our evaluations on an Ubuntu 20.04 machine with two AMD EPYC 7401 24-Core processors and \unit{256}{\giga\byte} RAM.
However, we ran a single-threaded version of our code and used just a few \giga\byte~of memory.
Thus, our algorithms have no expensive hardware requirements.
FF, EDF and Hermes were implemented in the same C++ framework as H2S and CELF.
We re-implemented GFH to add some small cache optimizations reducing the conflict graph building time and applying our system model. 
Python was used to build the CP model and solve it with CPLEX.
The implementations of H2S, CELF, FF, and EDF are single-threaded, while GFH and the CPLEX are designed for multithreaded execution, and used up to eight CPU cores.

\subsection{Scheduling with Queuing vs No-Wait Scheduling}

We started our evaluation on small communication scenarios with only six bridges and 450 streams, such that the CP solver can still find partial solutions within the time limit of \unit{2}{\hour}.
We excluded the Hermes algorithm from this scenario, because it is an all-or-nothing approach, i.e. Hermes either schedules all streams or none at all, and the evaluated stream set is oversaturating the network.
Therefore, Hermes would never return a schedule.

\begin{figure*}[]
	\centering
	\subfloat[Aggregated network throughput\label{fig:6_200:traffic}]{%
		\centering
		\includegraphics[width=0.33\linewidth]{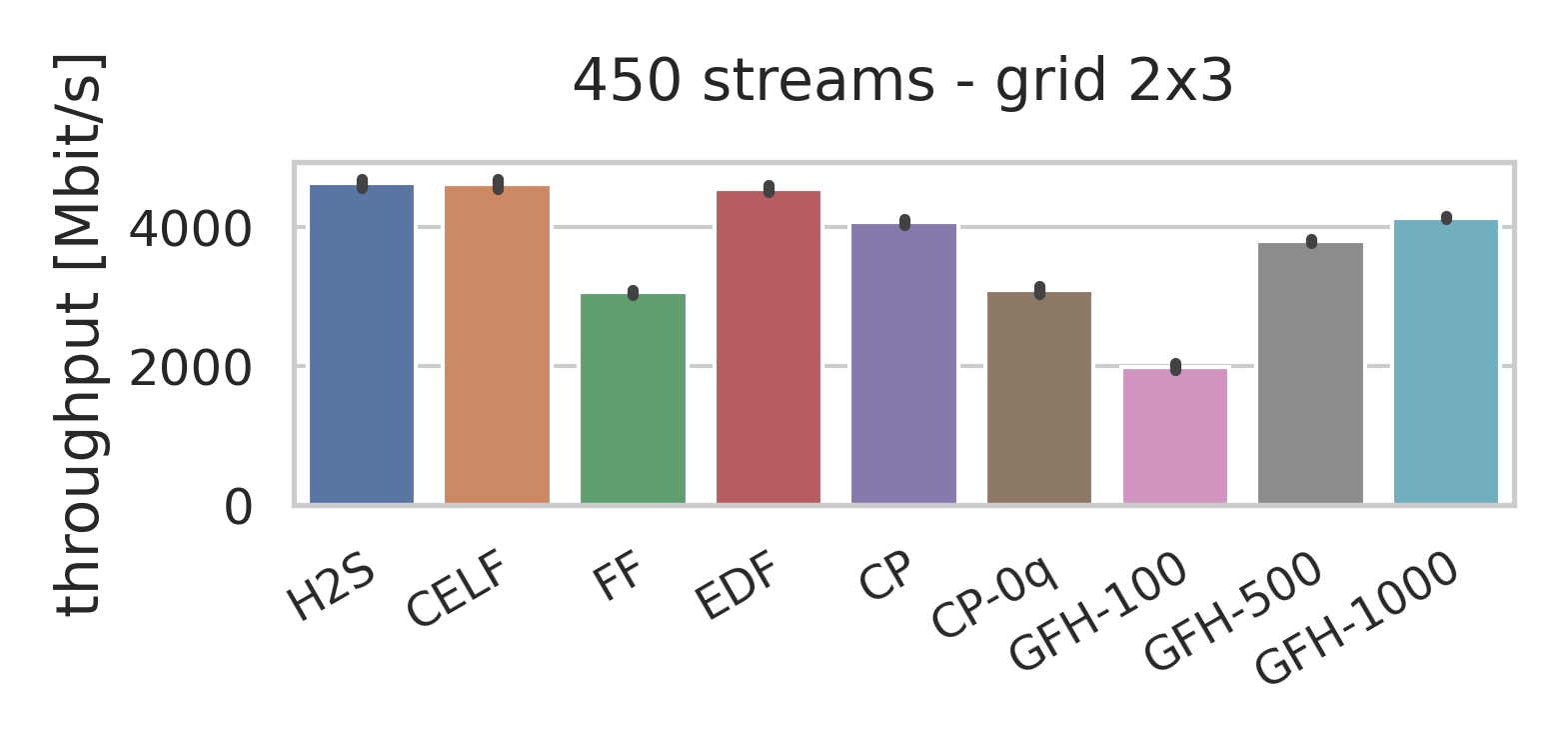}
	}
	~
	\subfloat[Admitted streams\label{fig:6_200:streams}]{%
		\centering
		\includegraphics[width=0.33\linewidth]{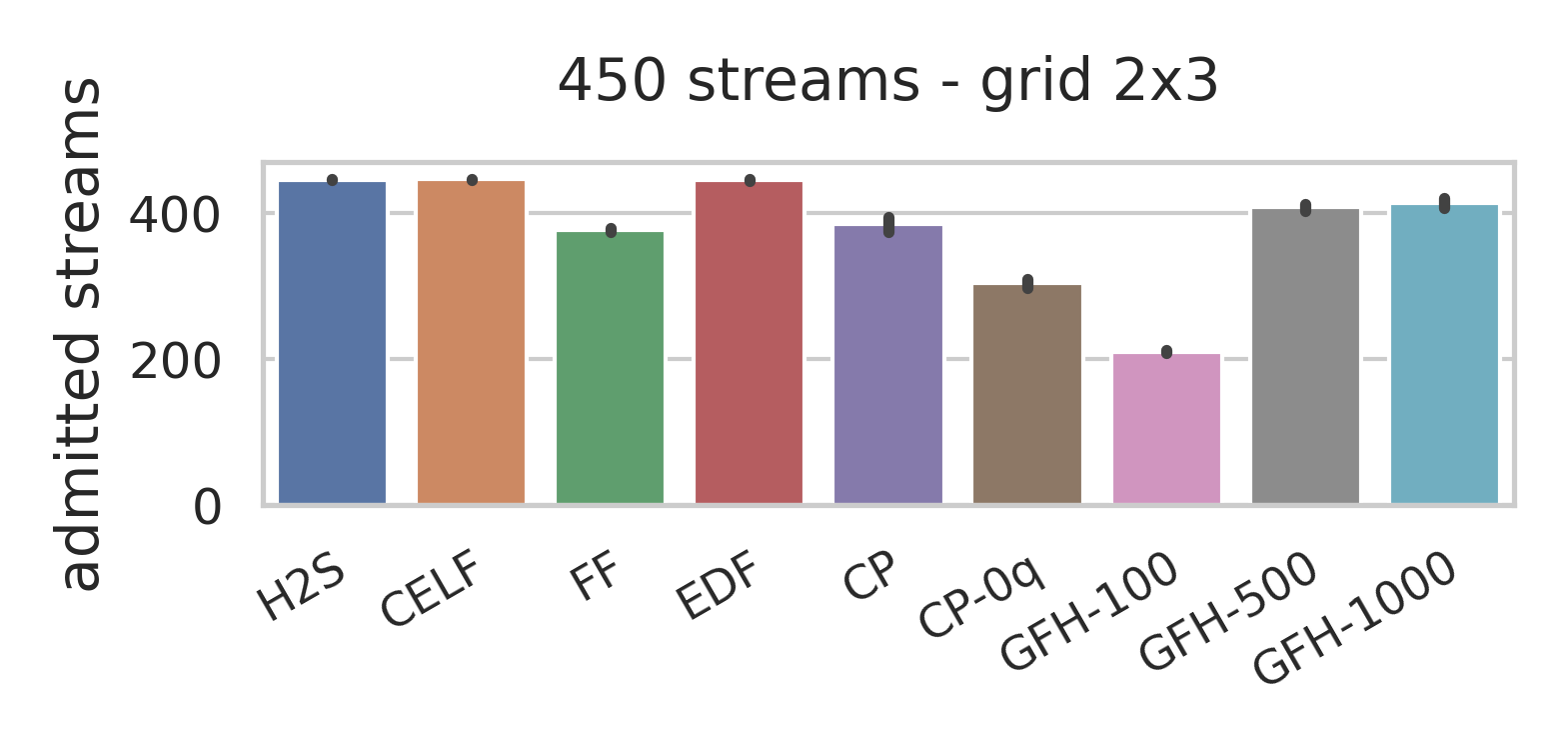}
	}
	~
	\subfloat[Solving time\label{fig:6_200:solving_time}]{%
		\centering
		\includegraphics[width=0.33\linewidth]{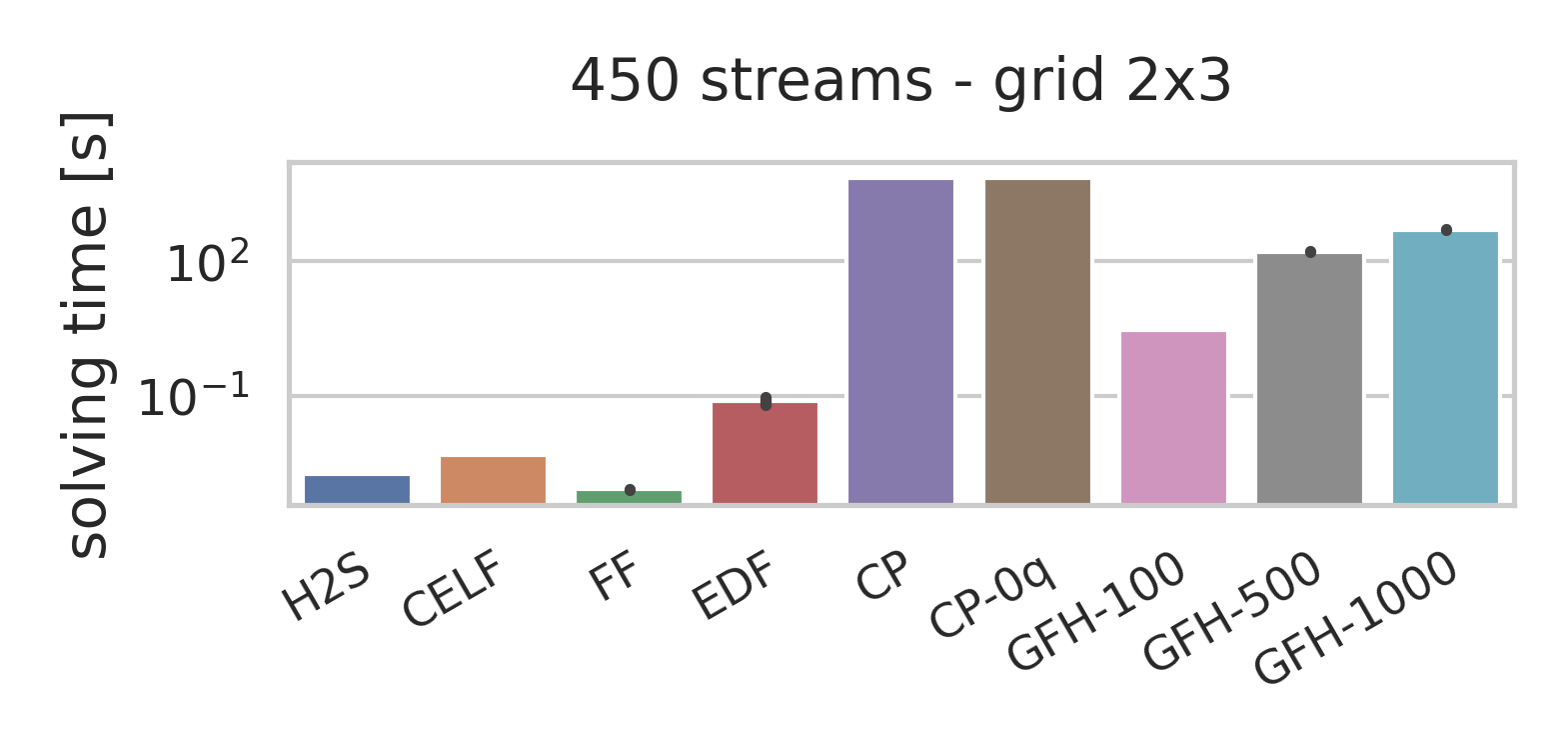}
	}
	\vspace{-0.2cm}
	\caption{Evaluation of 450 streams on an 2x3 grid network using various scheduling algorithms.}
	\label{fig:6_200}
	\vspace{-0.3cm}
\end{figure*}

H2S, CELF, and EDF all performed very well, yielding the highest aggregated throughput with no substantial difference (cf. Figure~\ref{fig:6_200:traffic}). 
In contrast, FF, CP, and GFH performed significantly worse.
Note that we ran GFH with different conflict graph sizes, but even with larger conflict graphs (GFH-500 and GFH-1000), the aggregated throughput was substantially lower than that of H2S, CELF, or EDF.
We ran the CP twice, once with queuing (CP) and once without queuing (CP-0q).
Here we again see the throughput difference between queuing and zero-queuing. 
Finally, we see that the trivial FF approach has difficulties finding good solutions with respect to the aggregated throughput in these small scenarios already.

In terms of admitted streams, the overall result looks quite similar.
Only FF improved slightly.
However, the results on admitted streams in Figure~\ref{fig:6_200:streams} are not that expressive, since the stream set is quite small and the well performing strategies admit almost all streams.
Hence, we can not see major trade-offs yet.
This will be discussed in more detail in Section~\ref{sec:eval:topology}.

We distinguish between the \emph{configuration time} as the time to create the internal representation of the problem in main memory (e.g. calculating candidate routes, setting up a conflict graph or creating a constraint set) and the \emph{solving time} as the time to run the actual algorithm on a pre-processed configuration.
With respect to solving time as shown in Figure~\ref{fig:6_200:solving_time}, we see that the CP solver hit the \unit{2}{\hour} time limit, while GFH ranged from seconds to minutes, depending on the conflict graph size.
The other heuristics were multiple orders of magnitude faster, finishing well under a second.
The configuration time was negligible for all algorithms except GFH (requiring more than \unit{40}{\minute} to build large conflict graphs, and still more than \unit{40}{\second} for small ones).

When looking at the solving time for the CP approach, it is apparent that even up-to-date CP solvers (or any other kind of solver) with modern hardware are not yet capable scaling to sufficiently large and complex scheduling problems.
A look at established solver based approaches shows that they often consider all-or-nothing problems, where the solver can either find a solution or state that it is unsolvable, e.g., \cite{Steiner2010}.
But when we ask for the best partial solution of a possibly unsolvable problem, the solver based approaches are still overstrained. 
Hence, we need fast heuristics to tackle these problems.

In our evaluation, we saw that the sophisticated queuing approaches, namely H2S, CELF, and EDF, performed better overall, resulting in schedules with an at least 20\% higher aggregated throughput.
This is because queuing allows for improved scheduling decisions.
Further, the complexity reduction of zero-queuing is not needed since the heuristics runtimes are very low already.

\subsection{Varying the Batch Sizes}
% Batch size eval

\begin{figure*}
	\subfloat[Final aggregated throughput\label{fig:batchsize:throughputall}]{%
		\includegraphics[width=0.30\linewidth]{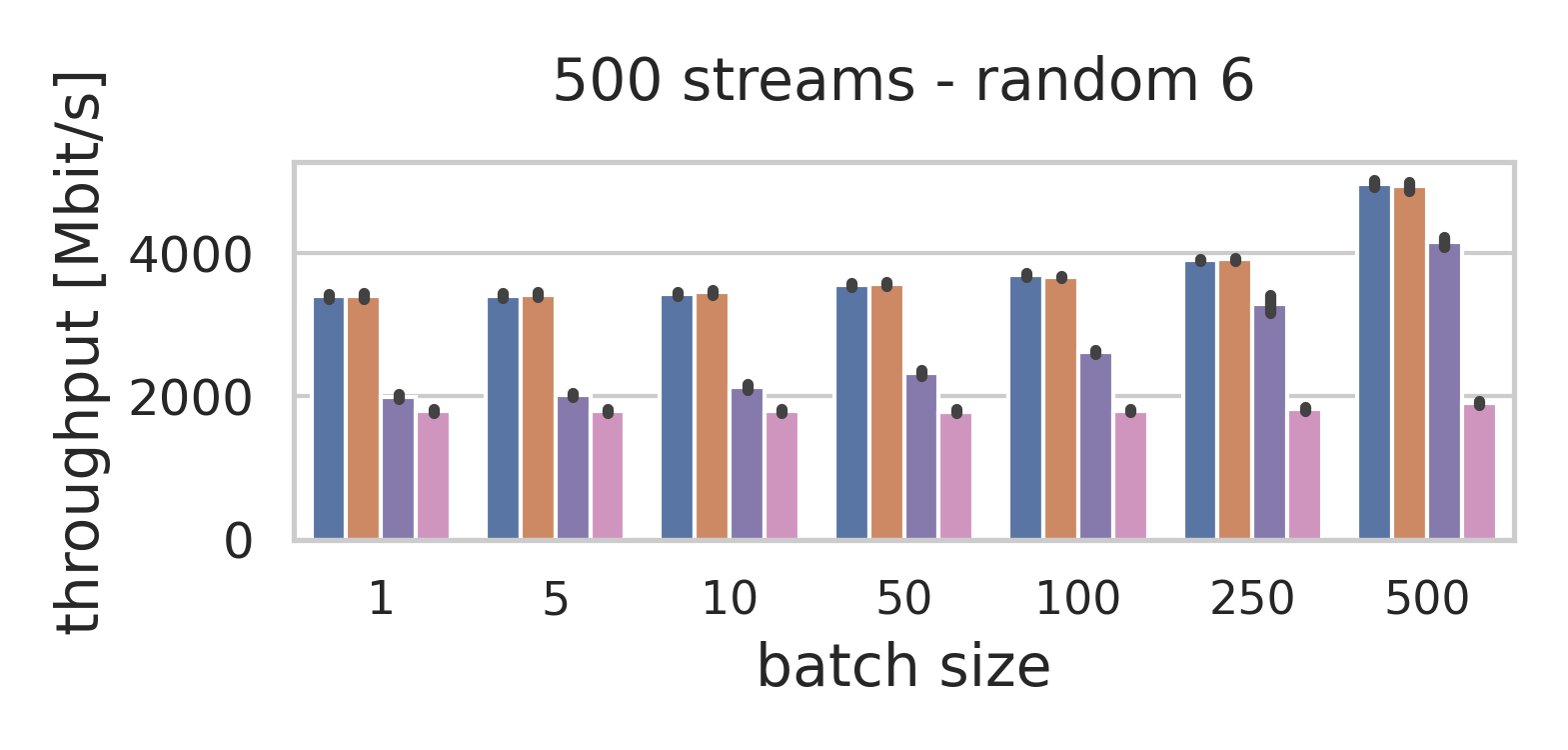}
	}%
	\subfloat[Solving time per batch\label{fig:batchsize:solvingtimeall}]{%
		\includegraphics[width=0.30\linewidth]{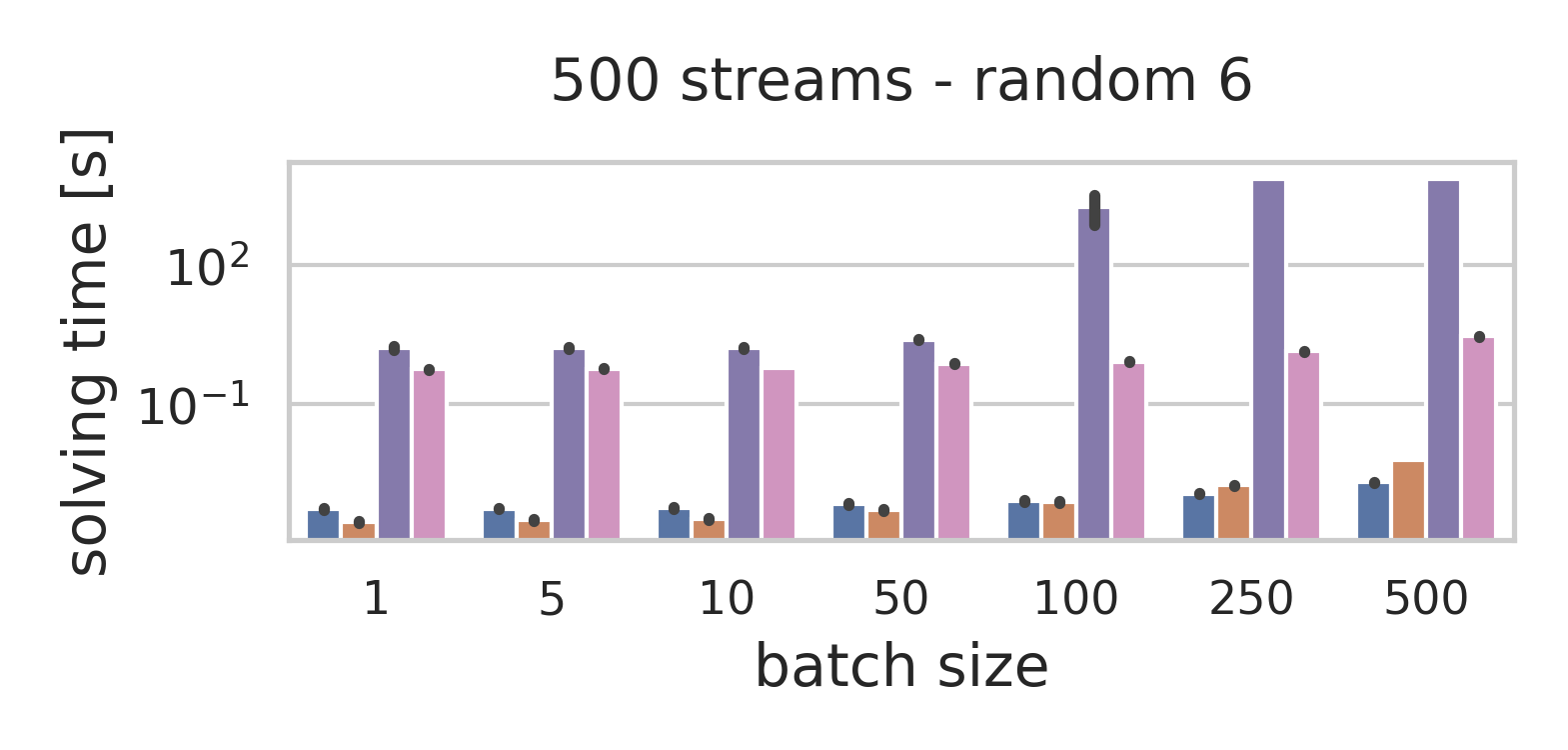}
	}%
	\subfloat[Cumulative solving time\label{fig:batchsize:cumulativesolvingtime}]{%
		\includegraphics[width=0.30\linewidth]{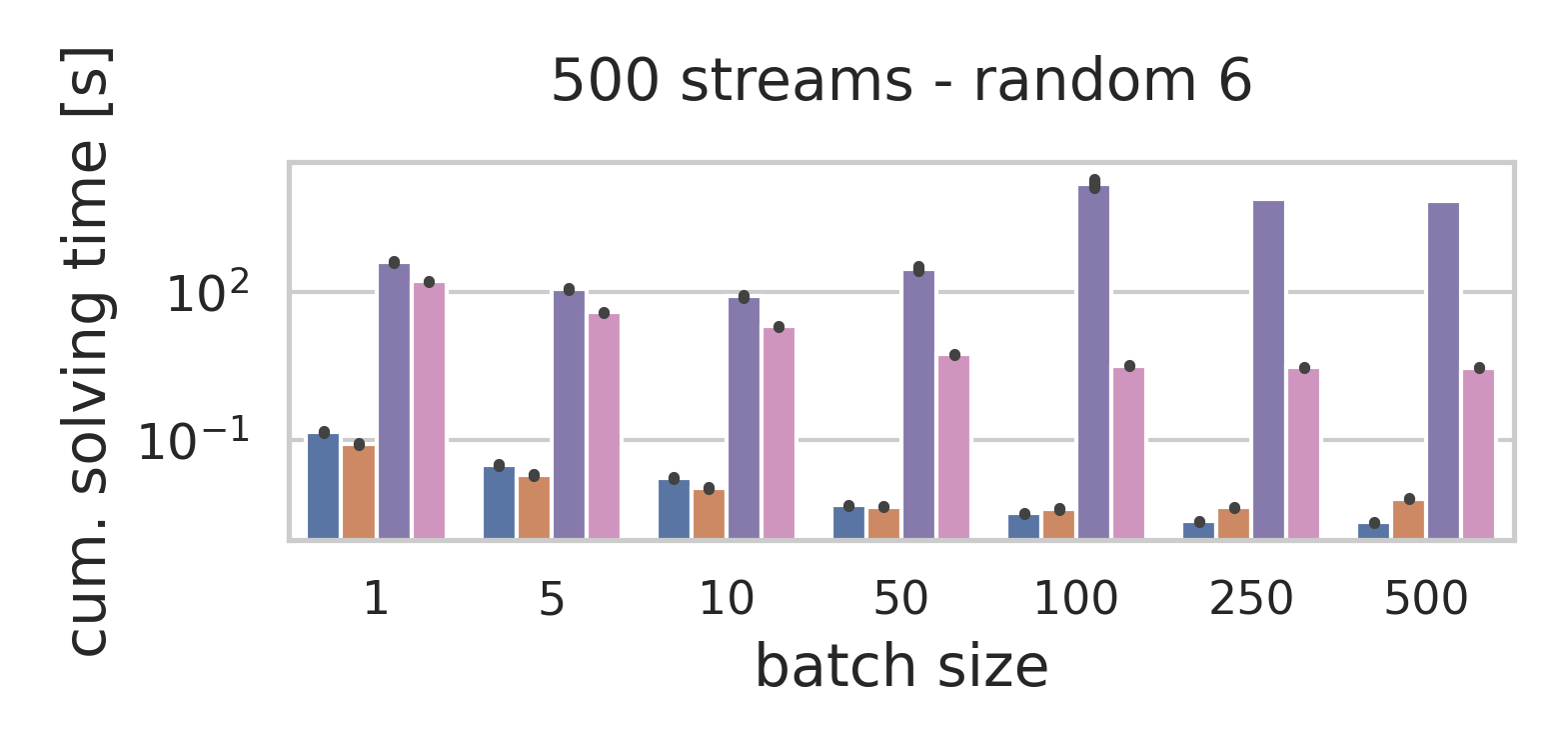}
	}%
	\subfloat[Legend\label{fig:batchsize:legend}]{%
		\includegraphics[width=0.09\linewidth]{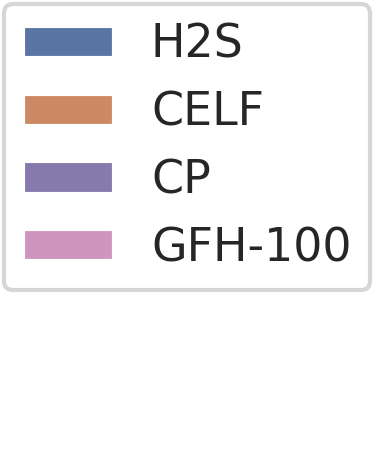}
		\vphantom{\includegraphics[width=0.2\textwidth]{plots/batch_size_scenario/bar_cumulative_solving_time_absolute-random-6-batch-size}}
	}%
	\caption{Evaluation of different batch sizes on random networks with 6 bridges and 500 streams.}
	\label{fig:batchsize}
\end{figure*}

Next, we took a closer look at the solving time and its impact on the overall average update latency, since the computation of new schedules obviously delays the roll-out of the schedules.
We evaluated different update batch sizes and present the aggregated throughput (Figure~\ref{fig:batchsize:throughputall}), the solving time per batch (Figure~\ref{fig:batchsize:solvingtimeall}), and the cumulative solving time of all batches until 500 streams have been handled by the scheduler (Figure~\ref{fig:batchsize:cumulativesolvingtime}).
Note that ``handled'' does not necessarily mean admitted.
In fact, for every batch size some streams were rejected.
The reason for the selected stream set size is that it shows the trade-off connected to the batch size and the resulting scheduling success.
We evaluated H2S and CELF against CP and GFH because they allow for defensive batch updates (GFH does offensive planning in some time steps, but most batches are scheduled defensively).
EDF, FF and Hermes were not evaluated, since Hermes would be overwhelmed by the oversaturation and EDF and FF use only update batches of size 1.

All scheduling strategies had lower solving times for small batches, but a larger cumulative solving time to compute all batches due to the additional overhead of every single batch.
At some point, CP showed a different behavior when single batches start reaching the \unit{2}{\hour} time limit.
The total solving time was then getting smaller because there are fewer batches to hit the limit.

The aggregated throughput was much better (30--50\%) for H2S, CELF, and CP when very large batches are used, i.e. in offline planning.
We did not see a significant difference between H2S and CELF.
GFH's aggregated throughput was almost constant, decreasing slightly with larger batch sizes.
However, we did not observe this behavior on larger conflict graphs.
The constant behavior comes from GFH's offensive planning which was triggered several times, thus fixing suboptimal choices from previous batches.
GFH's aggregated throughput increased for larger conflict graphs, as did the solving and configuration time.

In summary, when the acceptable update latency needs to be low and the average request rate is also low, i.e. the batch sizes are small, H2S and CELF provided small average update latencies and still offered decent aggregated throughput.
However, H2S and CELF do not need small batch sizes, since they are fast enough already.

\subsection{Comparison Against a State-of-the-Art Queuing Approach}

We compared H2S and CELF to Hermes on multiple tree networks with 15 bridges and 15 end devices each.
Therefore, we looked at scenarios with 200--300 streams and 50 samples for each scenario. 
The tree topologies were used to ensure that Hermes has no deadlocks, which can arise when the stream routes form a circle, because Hermes schedules on a per network link basis.
Further, we deliberately limited the number of candidate paths to one for H2S and CELF by using tree networks, since Hermes does not support multiple candidate paths.
Because Hermes is an all-or-nothing scheduling algorithm, admitting either every stream or none, we present the schedulability, i.e., the portion of scenarios where a strategy admitted all streams, instead of the aggregated throughput. 

\begin{figure}[]
	\centering
		\includegraphics[width=0.7\linewidth]{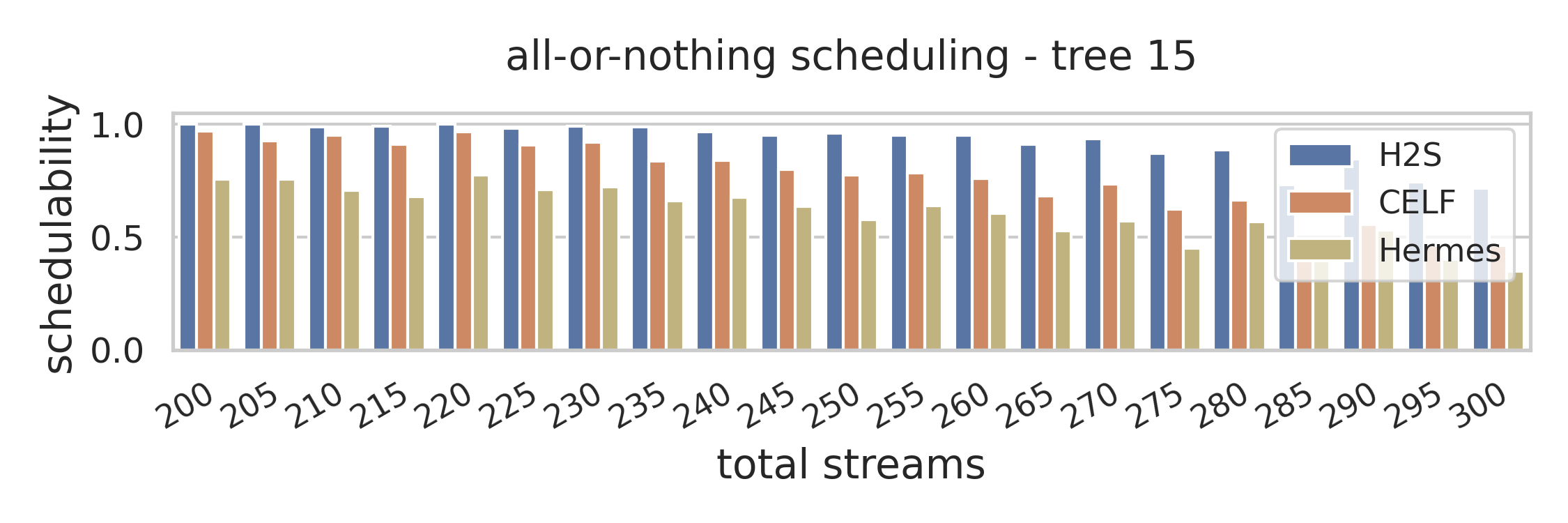}
	\vspace{-0.2cm}
	\caption{Success rate in an \emph{All-or-nothing} scenario evaluation with 200-300 streams on different trees with 15 bridges}
	\label{fig:15_all_or_noting}
\end{figure}

H2S consistently outperformed CELF and Hermes (cf. Figure~\ref{fig:15_all_or_noting}).
CELF reported a higher schedulability than Hermes in most cases, but not always substantially.
The poor performance of Hermes can be explained by its network link based scheduling.
Hermes first determines precedence constraints from the stream routes for every link and then schedules all forwarding operations link by link.
This way, streams can be delayed due to chained precedence constraints, which in turn can lead to deadline misses.
The CELF adaptation often performed similar to H2S in terms of rejected streams and throughput.
But, apparently, CELF had to reject at least one stream in much more samples than H2S, what lead to significant differences in the schedulability.
This might be subject to the network utilization element in the CELF rating function (cf.~\ref{eq:crf}), where longer routes are scheduled later and by then might be unplaceable. 
Considering the runtime, H2S and CELF were almost one order of magnitude faster than Hermes.
However, all strategies finished within a fraction of a second, making the difference negligible.

\subsection{Evaluation on Different Network Topologies}
\label{sec:eval:topology}

\begin{figure*}
	\subfloat[Throughput on random topologies\label{fig:bartrafficabsolute-random-25-scenario25hosts-2}]{%
		\centering
		\includegraphics[width=0.30\linewidth]{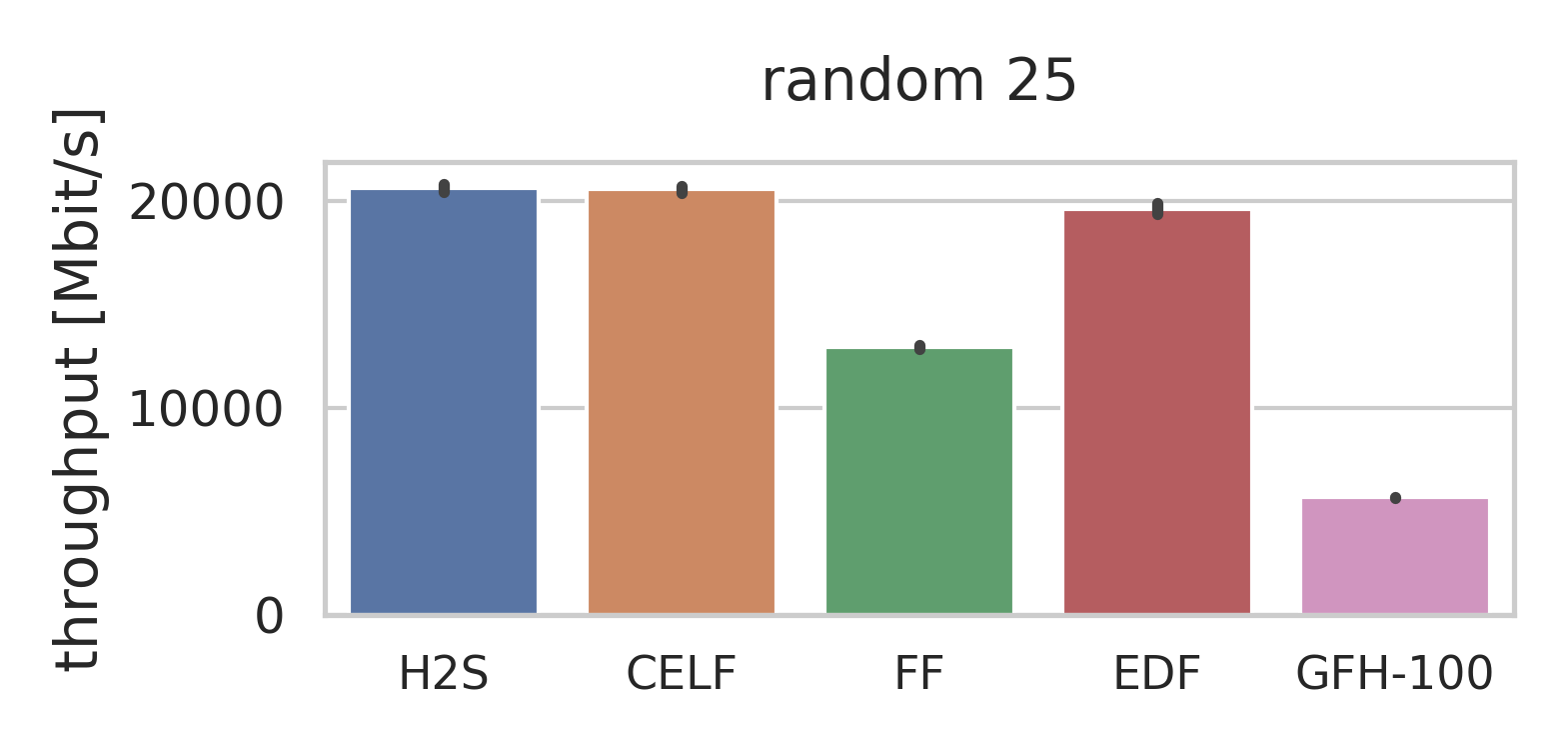}
	}
	\qquad
	\subfloat[Throughput on ring topologies\label{fig:bartrafficabsolute-ring-25-scenario25hosts-2}]{%
		\centering
		\includegraphics[width=0.30\linewidth]{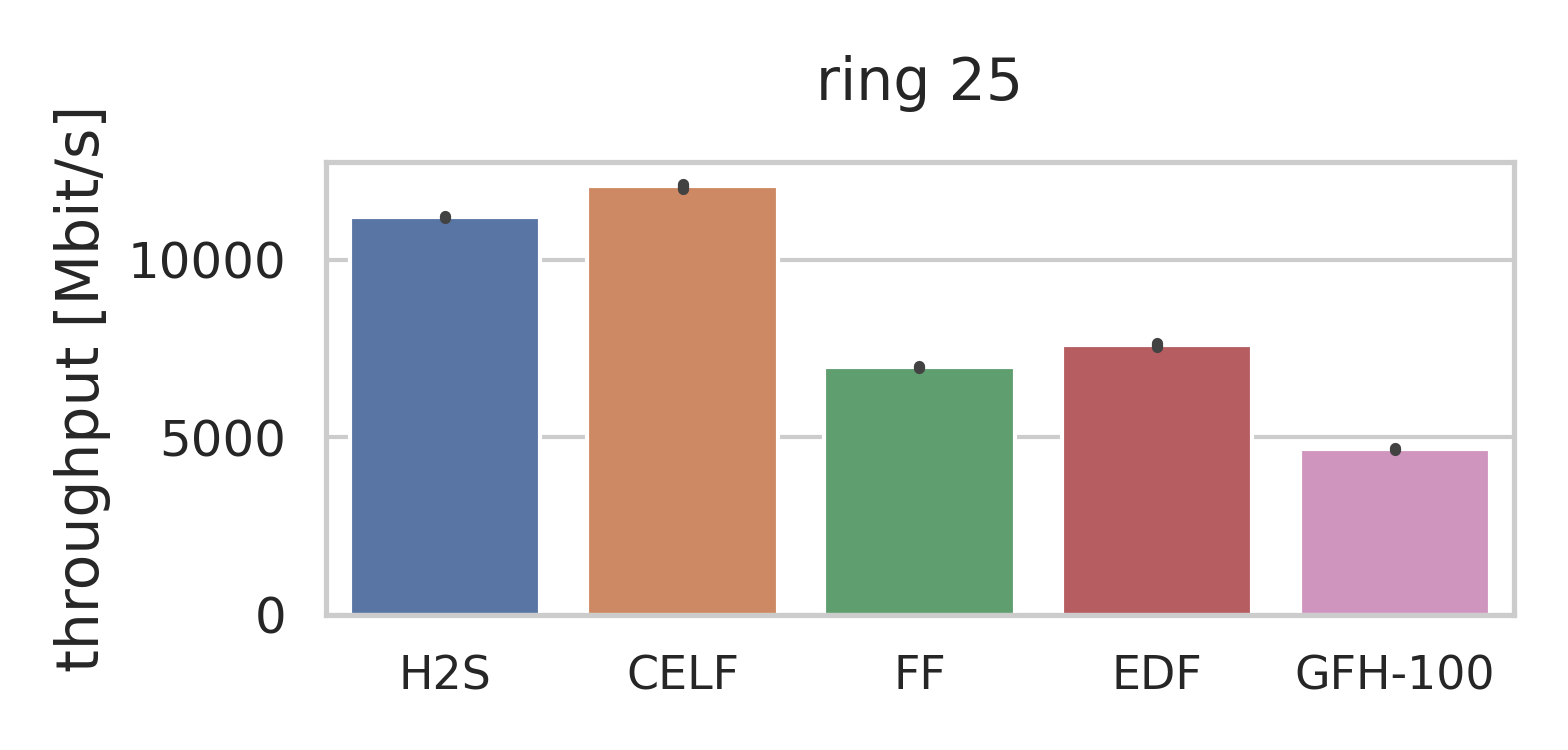}
	}
	\qquad
	\subfloat[Throughput on tree topologies\label{fig:bartrafficabsolute-tree-25-scenario25hosts-2}]{%
		\includegraphics[width=0.30\linewidth]{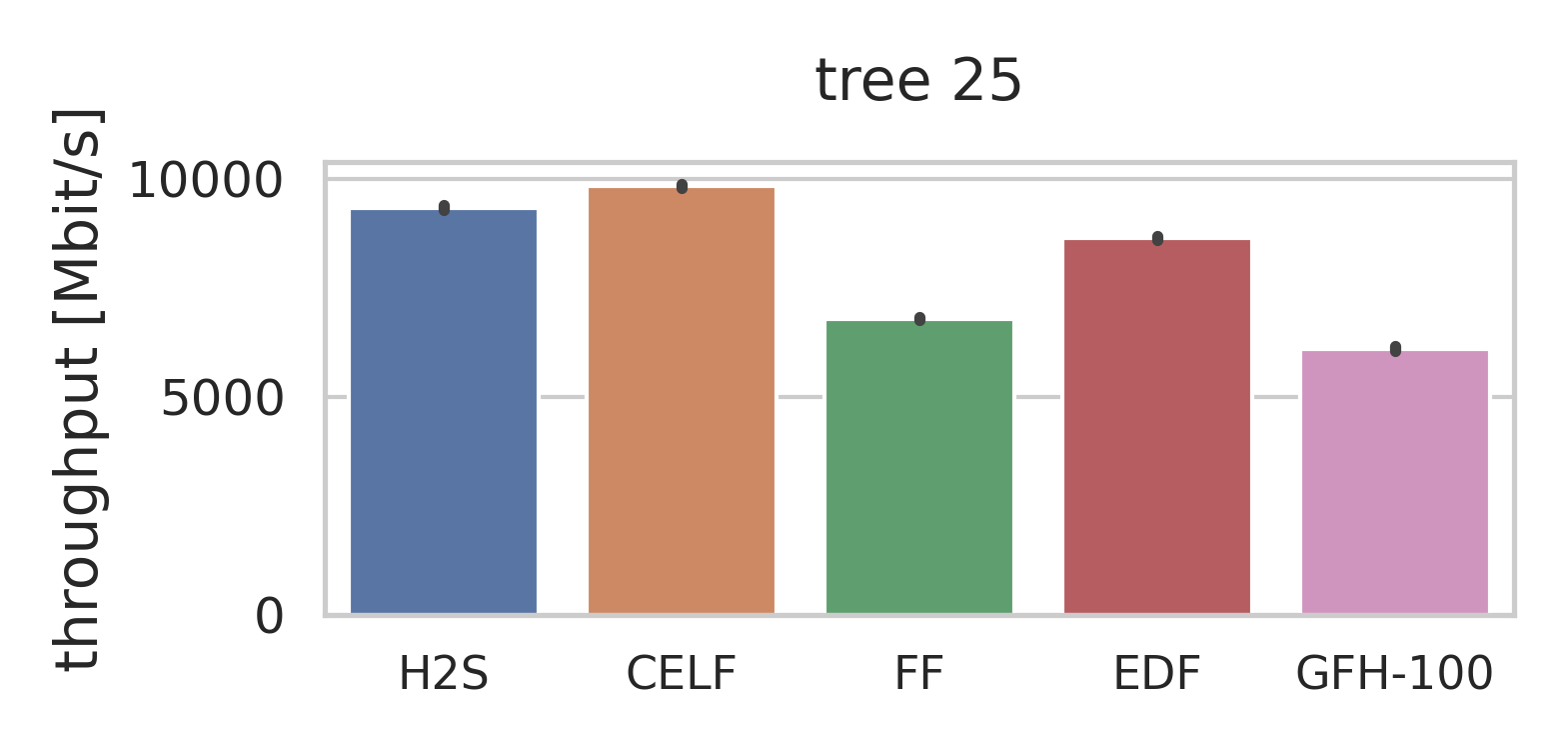}
	}
\\
	\subfloat[Accepted streams on random topologies\label{fig:baracceptedflowsabs-random-25-scenario25hosts-2}]{%
		\includegraphics[width=0.30\linewidth]{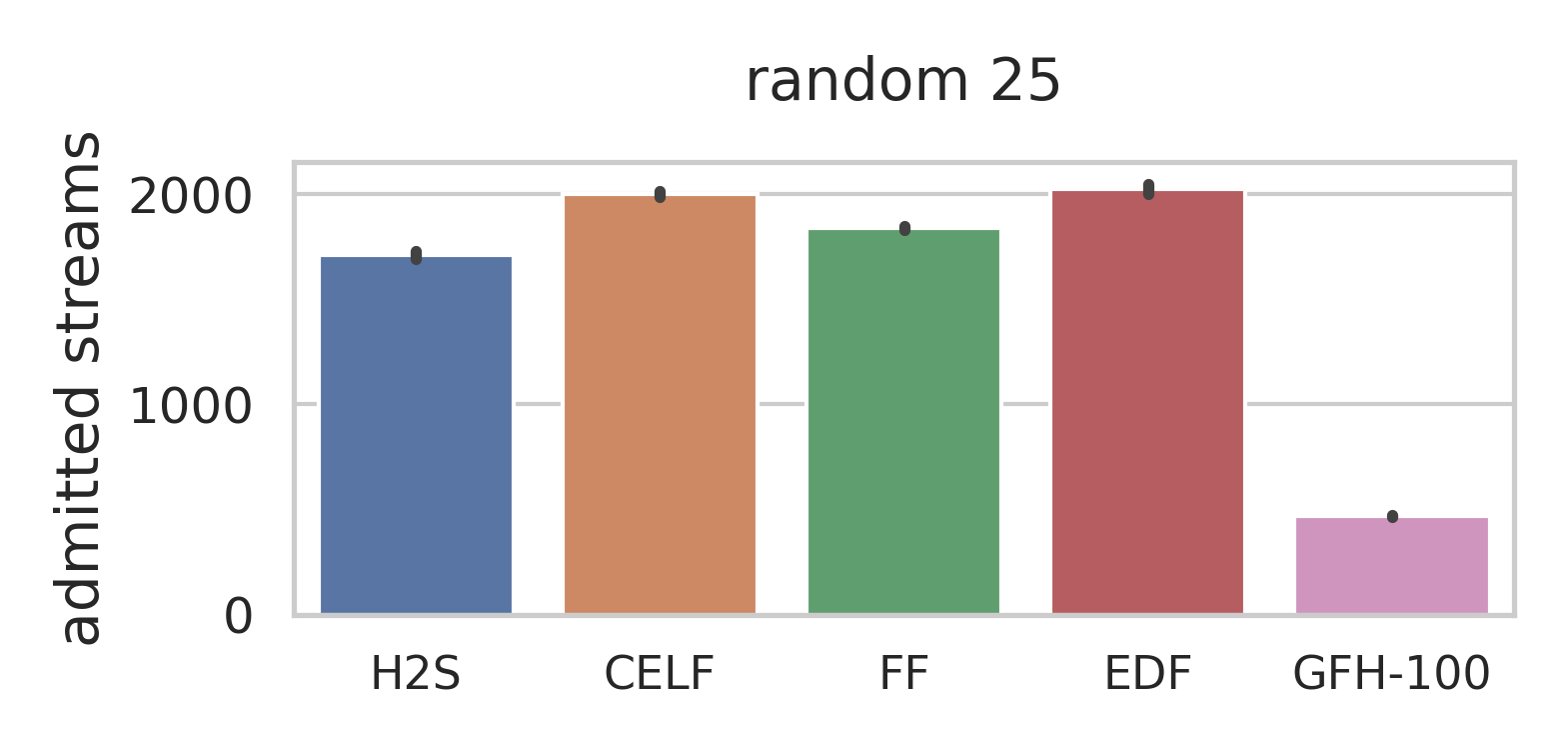}
	}
	\qquad
	\subfloat[Accepted streams on ring topologies\label{fig:baracceptedflowsabs-ring-25-scenario25hosts-2}]{%
		\includegraphics[width=0.30\linewidth]{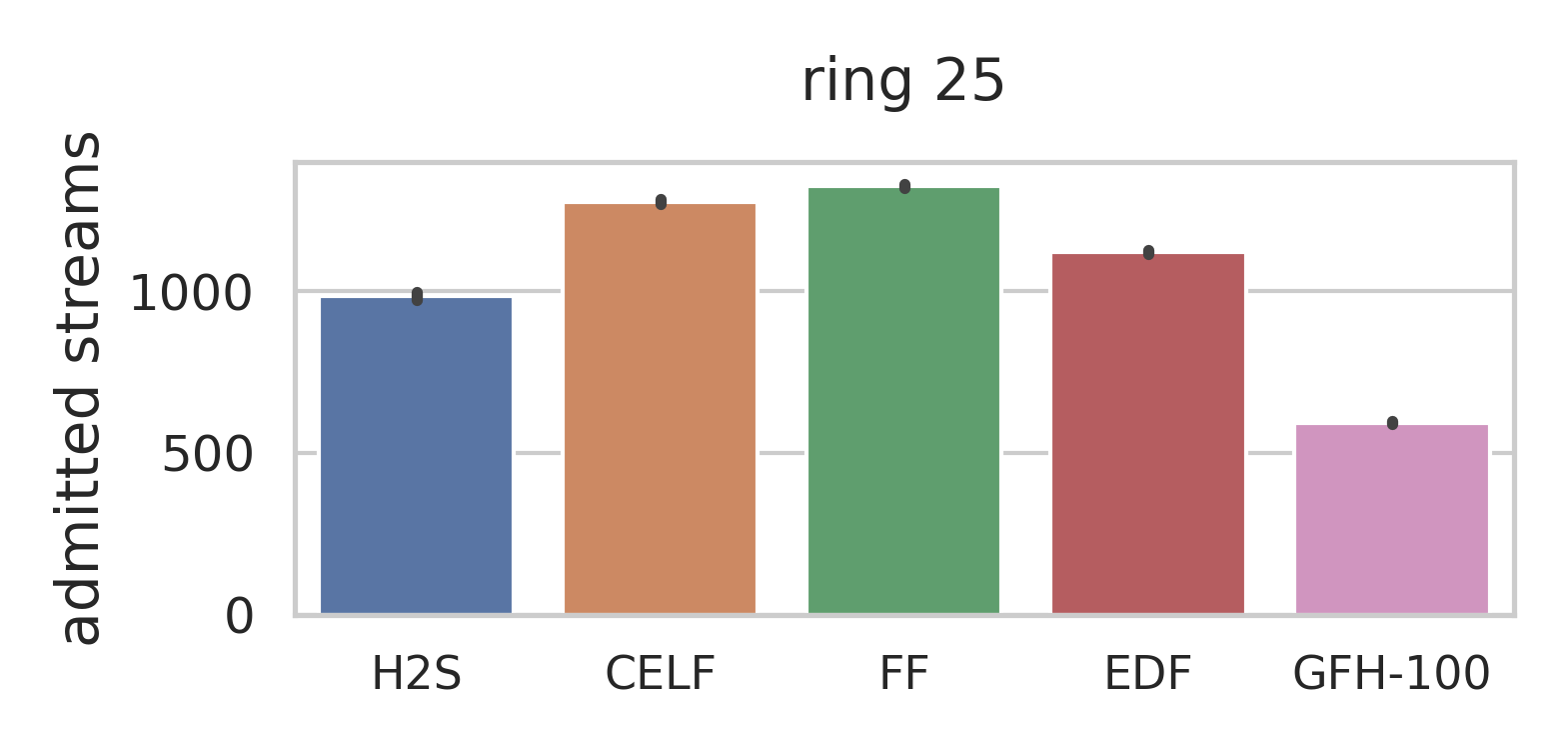}
	}
	\qquad
	\subfloat[Accepted streams on tree topologies\label{fig:baracceptedflowsabs-tree-25-scenario25hosts-2}]{%
		\includegraphics[width=0.30\linewidth]{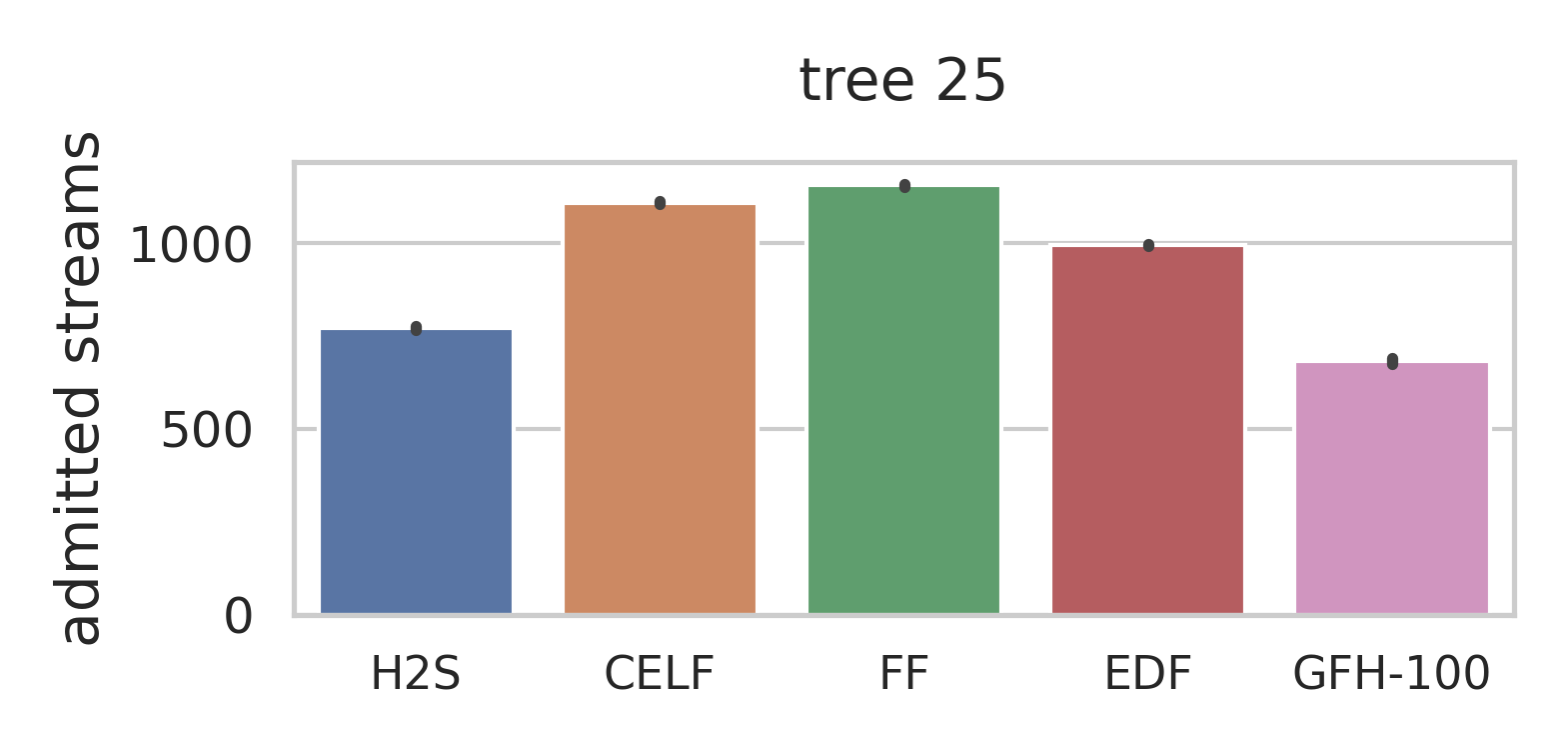}
	}
\\
	\subfloat[Runtime on random topologies\label{fig:barsolvingtimeabsolute-random-25-scenario25hosts-2}]{%
		\includegraphics[width=0.30\linewidth]{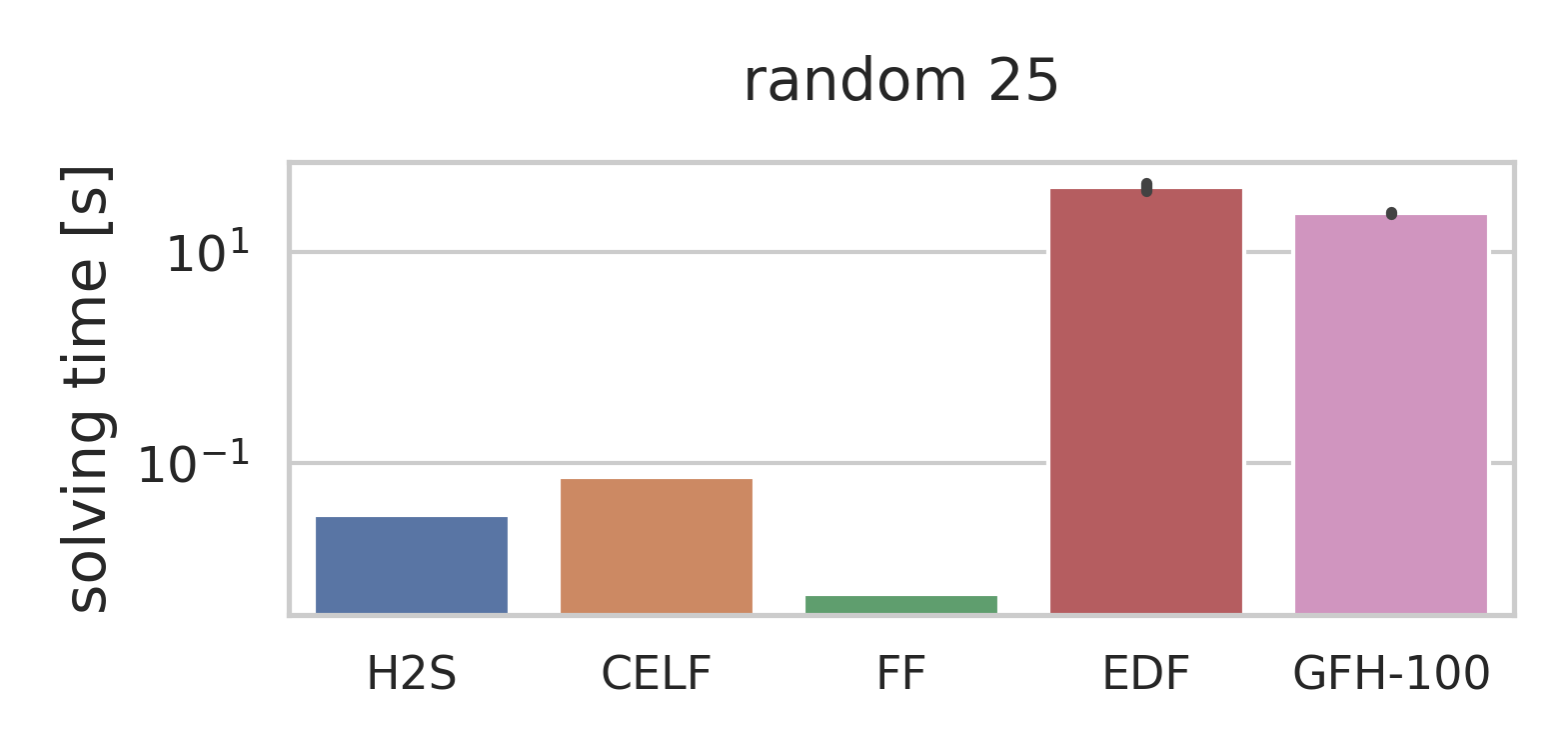}
	}
	\qquad
	\subfloat[Runtime on ring topologies\label{fig:barsolvingtimeabsolute-ring-25-scenario25hosts-2}]{%
		\includegraphics[width=0.30\linewidth]{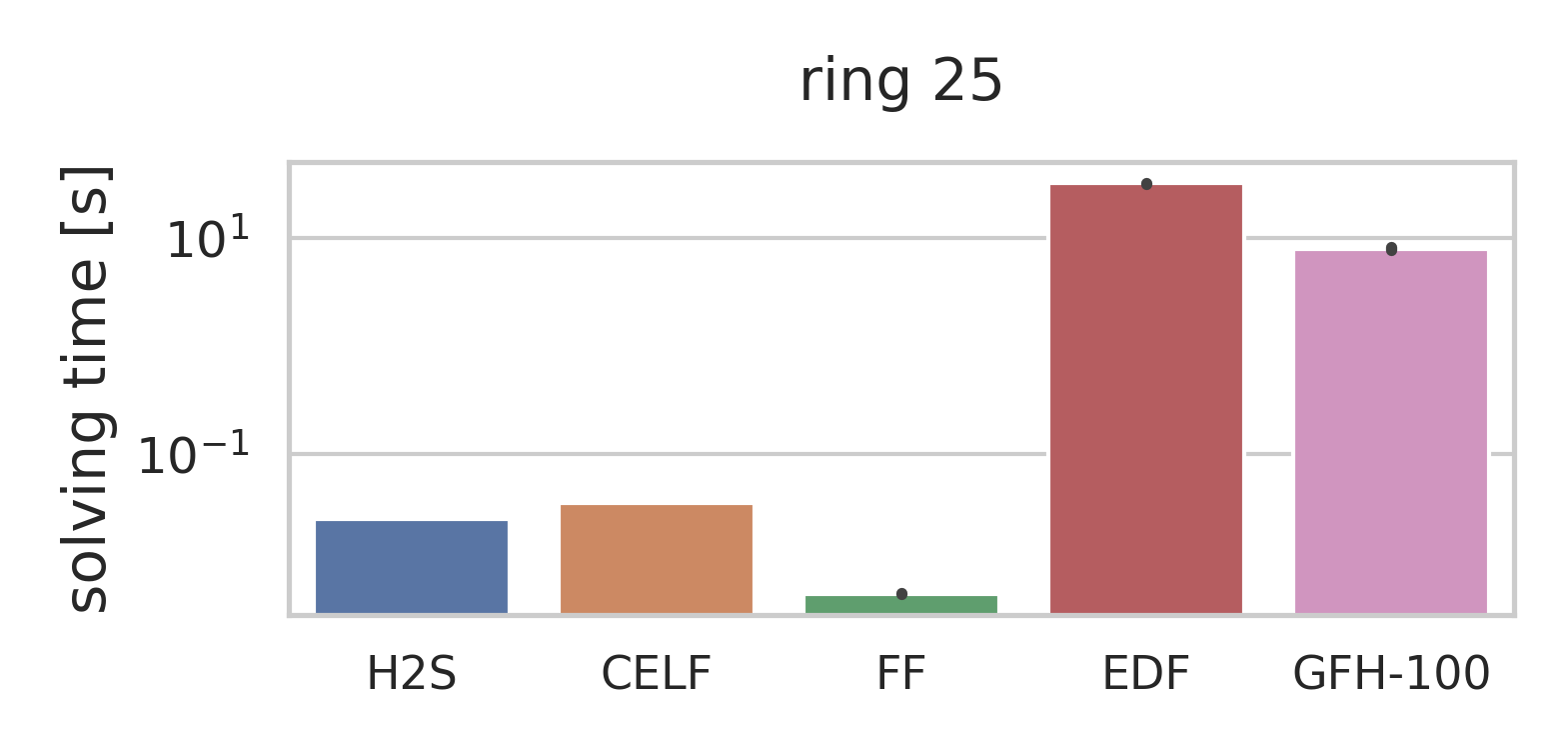}
	}
	\qquad
	\subfloat[Runtime on tree topologies\label{fig:barsolvingtimeabsolute-tree-25-scenario25hosts-2}]{%
		\includegraphics[width=0.30\linewidth]{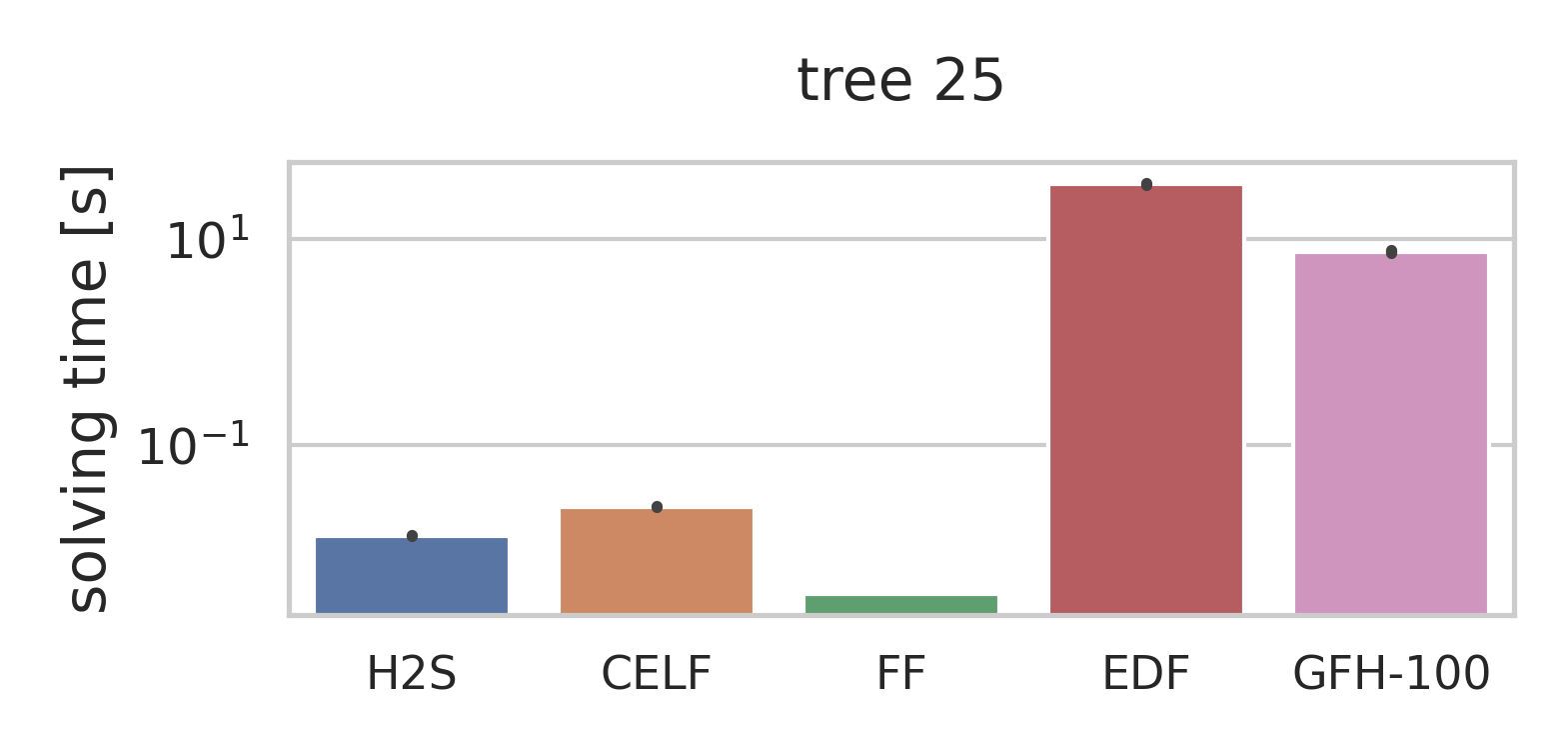}
	}
	\caption{Topology evaluation on networks with 25 bridges}
	\label{fig:topology}
\end{figure*}

% environment
We evaluated H2S and CELF alongside FF, EDF, and GFH (with 100 configurations per stream) on networks with 25 bridges arranged in an Erd\"os-R\'enyi random, grid, ring, and tree structure to observe the impact of the underlying network topology.
Thereby, we tried to schedule 2500 streams, which is much more than any of the tested strategies was able to schedule on networks of size 25.
This way, we see the divergence in the number of admitted streams and aggregated throughput.
CP was excluded because it takes too long (weeks per sample) and Hermes was left out since it again would not yield any schedules due to the oversaturation. 

% results
The results are displayed in Figure~\ref{fig:topology}. 
We omitted the grid topology plots since the algorithms behaved similarly to the random networks.
Overall, H2S and CELF performed very well, having the highest aggregated throughput and a solving time always below \unit{0.04}{\second} and \unit{0.08}{\second}, respectively.
Meanwhile, GFH took up to \unit{26}{\second} of solving time and additionally \unit{12-30}{\minute} to build the conflict graph.
A notable result was the difference in the number of admitted streams of H2S and CELF on the random network while having a similar aggregated throughput.
CELF scheduled more streams with lower aggregated throughput per stream.
Note that the number of scheduled streams is not necessarily a meaningful metric, because scheduling many high period streams can be much simpler than scheduling a few low period streams.
Therefore, the aggregated network throughput is a much better indicator for the scheduling quality.
In the case of the ring and tree topology, CELF performed better than H2S by scheduling more streams and having a higher throughput, since CELF avoids stream placements along very suboptimal routes (e.g. taking the long way in a ring).
Meanwhile, EDF yielded results comparable to CELF on the random topology, but falls behind on the ring and tree and throughout has much higher solving times of up to \unit{1}{\minute}.
FF and GFH were never competitive in terms of throughput.

Overall, all algorithms showed significant differences in terms of the number of accepted streams, as well as in the aggregated throughput, when comparing their results from different network topologies.
This can be explained by the much smaller diameter in the random network, compared to the other two and the existence of multiple feasible candidate routes.
In the ring and tree networks, many more streams must share bottleneck links that cannot be bypassed.

\subsection{Large-scale Network Evaluation}
% large-scale evaluation

Next, we scaled the networks to 1,000 bridges and 48k streams.
Note that the other scheduling algorithms do not scale well enough for such large networks and were therefore excluded.
H2S and CELF kept outperforming FF in terms of aggregated throughput and admitted streams (cf. Table~\ref{tab:large_scale_eval}).
The only exception was on the large tree network. 
Here, the topology limited the admitted streams significantly, so that FF could catch up with H2S in terms of admitted streams.
However, H2S and CELF had almost twice the aggregated throughput than FF.
CELF did not pay off in the large networks, yielding a slightly lower aggregated throughput than H2S.
However, CELF still provided a higher number of admitted streams, thus offering a different trade-off between aggregated throughput and admitted streams.
The significant difference in the throughput numbers between the different networks was attributable to the distinctive topologies and their characteristics, e.g., average candidate route length and number of possible candidate routes.

% solving time
In terms of solving time, FF was the fastest strategy, since the heuristic performs almost no computation.
However, H2S scheduled random and tree networks within less than a second and the grid within \unit{3}{\second}.
CELF always took a second or more, yet still had very short solving times.
Thus, H2S and CELF scale well to this enormous network and stream set sizes.

\begin{table*}[t]
\centering
\small
\renewcommand{\arraystretch}{1.1}
\begin{tabular}{|c|p{1.5cm}|p{1.5cm}|p{1.5cm}|p{1.5cm}|p{2.25cm}|p{2cm}|}
	\hline
	\multirow{1}{*}{topology} & strategy & throughput \newline [\giga\bit\per\second] & admitted \newline streams & solving \newline time [\second] & mean scheduling \newline table length & max scheduling \newline table length \\
	\hline
	\multirow{3}{*}{random 1000} 
	& H2S & 494.681 & 47,937 & 0.821 & 165 & 430 \\
	\cline{2-7}
	& CELF & 493.929 & 47,952 & 3.529 & 168 & 435 \\
	\cline{2-7}
	& FF & 335.712 & 31,494 & 0.262 & 118 & 296 \\
	\hline
	\multirow{3}{*}{grid 40x25} 
	& H2S & 202.338 & 27,208 & 3.031 & 258 & 683 \\
	\cline{2-7}
	& CELF & 192.291 & 28,560 & 5.708 & 273 & 737 \\
	\cline{2-7}
	& FF & 113.372 & 24,037 & 0.364 & 158 & 467 \\
	\hline
	\multirow{3}{*}{tree 1000} 
	& H2S & 47.876 & 5,443 & 0.773 & 72 & 684 \\
	\cline{2-7}
	& CELF & 47.444 & 6,312 & 1.186 & 83 & 833 \\
	\cline{2-7}
	& FF & 24.381 & 5,518 & 0.112 & 48 & 580 \\
	\hline
\end{tabular}
\caption{Large-scale evaluation with 48k streams on networks with 1,000 bridges}
\label{tab:large_scale_eval}
\end{table*}

% scheduling table size
In Table~\ref{tab:large_scale_eval}, we also give an indication on the resulting scheduling table sizes.
In our case, this represents the number of frames forwarded, since we compute precise per frame forwarding operations.
Thereby, we see a significant difference between the mean scheduling table sizes and the maximum size, i.e., the longest table in the resulting schedule.
This is because some network links are used much more.
In terms of feasibility, the table lengths are within the order of magnitude of modern local area network switches.

To summarize, we saw that H2S and CELF scaled to very large networks and huge stream sets, while the aggregated throughput and the number of admitted streams was still significantly larger than the baseline algorithm (FF).
The solving time was still in the area of sub-seconds to seconds and the trade-offs that H2S and CELF offer stay intact.

\subsection{Dynamic Systems}

\begin{figure*}
	\subfloat[Throughput on random topologies\label{fig:dynamic_scenario:traffic-random}]{%
		\centering
		\includegraphics[width=0.31\linewidth]{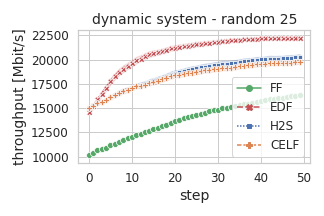}
	}
	~
	\subfloat[Accepted streams on random topologies\label{fig:dynamic_scenario:accepted_streams-random}]{%
		\centering
		\includegraphics[width=0.31\linewidth]{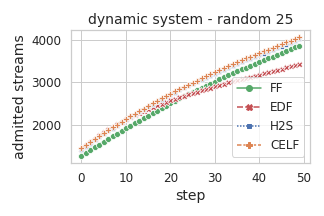}
	}
	~
	\subfloat[Solving time on random topologies\label{fig:dynamic_scenario:solving_time-random}]{%
		\includegraphics[width=0.31\linewidth]{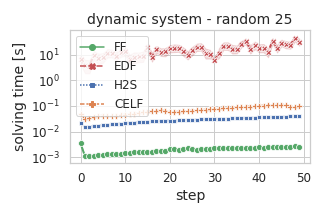}
	}
	\\
	\subfloat[Throughput on ring topologies\label{fig:dynamic_scenario:traffic-ring}]{%
		\includegraphics[width=0.31\linewidth]{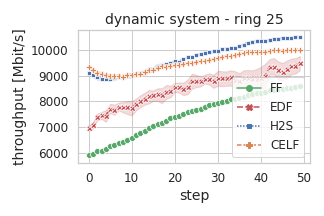}
	}
	~
	\subfloat[Accepted streams on ring topologies\label{fig:dynamic_scenario:accepted_streams-ring}]{%
		\includegraphics[width=0.31\linewidth]{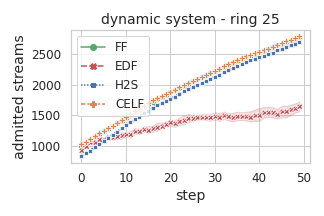}
	}
	~
	\subfloat[Solving time on ring topologies\label{fig:dynamic_scenario:solving_time-ring}]{%
		\includegraphics[width=0.31\linewidth]{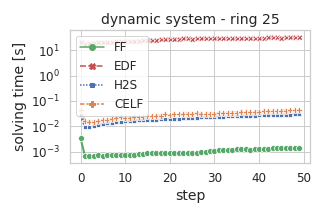}
	}
	\caption{Topology evaluation on networks with 25 bridges}
	\label{fig:dynamic_scenario}
\end{figure*}

Finally, we took a look at a dynamic system, where streams leave, and new ones appear over time.
We tried to schedule 1500 streams in the first step, followed by 50 time steps where 100 streams left and 200 streams entered the system in every time step.
Thus, the network become more congested over time while at least some resources become available.
As a result, handling the emerging fragmentation becomes the main challenge.
Again, we used the networks with 25 bridges, and provide the results for the random and ring topologies in Figure~\ref{fig:dynamic_scenario}. 
The results for the grid network looked similar to the random topology, and the tree topology data being in between the ring and random network results.

In this evaluation we compared H2S, CELF, EDF and FF.
The CP and GFH did not scale sufficiently to allow for fast scheduling adaptations.
Hermes would run into deadlocks, unable to return valid schedules.

When comparing the aggregated network throughput, we saw that EDF yielded the best results for the random network, since it is an offensive planning strategy by design.
However, H2S and CELF provided good results, improving substantially upon the borderline FF quality, and they were capable of handling the fragmentation issue.
This is especially visible on the ring topology (cf. Figure~\ref{fig:dynamic_scenario:traffic-ring}), where H2S and CELF started very well, but then the throughput decreased for a few time steps.
The reason for this was that there were not sufficient gaps in the schedule to reinsert the new streams.
However, after several time steps, the fragmentation become better and both strategies improve beyond the initial throughput.
Further, H2S and CELF outperformed EDF on the ring network.
Yet, EDF caught up over time.

Turning to the evaluation of the accepted streams, we saw that EDF admits fewer streams over time, especially in the ring network.
This was because the high traffic streams usually also have short periods and thus tight deadlines, which are preferred by EDF.
Thus, many high period streams were rejected.
The other strategies all have a comparable number of admitted streams which sometimes made it almost impossible to see the data points. 
When looking at the solving time, we saw the same trends as before.
EDF took several orders of magnitude longer to yield a schedule (more than \unit{10}{\second}), while H2S, CELF and FF finished within fractions of a second.
The very first solving time is relatively high for H2S, CELF and FF, since in step 0, they had to schedule 1500 streams, in contrast to only 200 streams afterwards.

In summary, we saw that the offensive planning mode of H2S and CELF is fast and scalable, outperforming FF substantially in terms of aggregated throughput.
EDF, which was designed as offensive strategy, uses more degrees of freedom, and takes much more time to compute schedules achieves higher throughput.
However, EDF has starting issues and needed some time until it overtook H2S and CELF in terms of throughput.
Especially considering the runtime of H2S and CELF, they are good options to dynamically compute new schedules.

\subsection{Case-study Energy System}

\begin{figure*}
	\subfloat[H2S\label{fig:ami:success:H2S}]{%
		\includegraphics[width=0.24\linewidth]{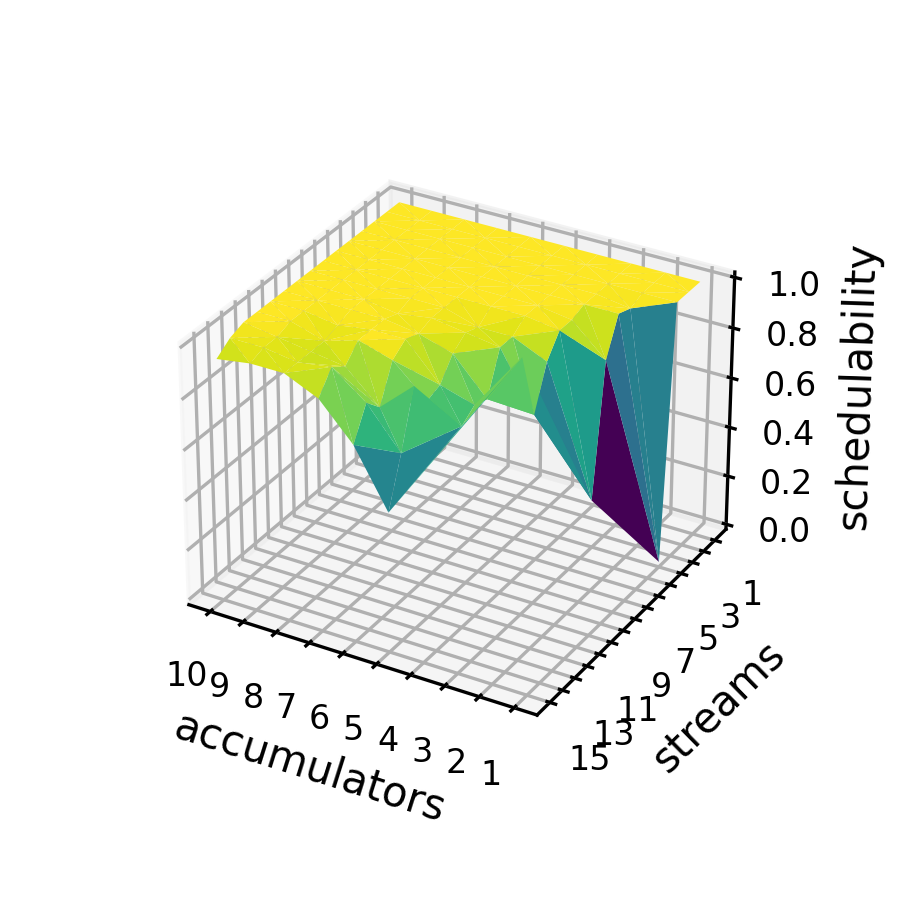}
	}%
	\subfloat[CELF\label{fig:ami:success:CELF}]{%
		\includegraphics[width=0.24\linewidth]{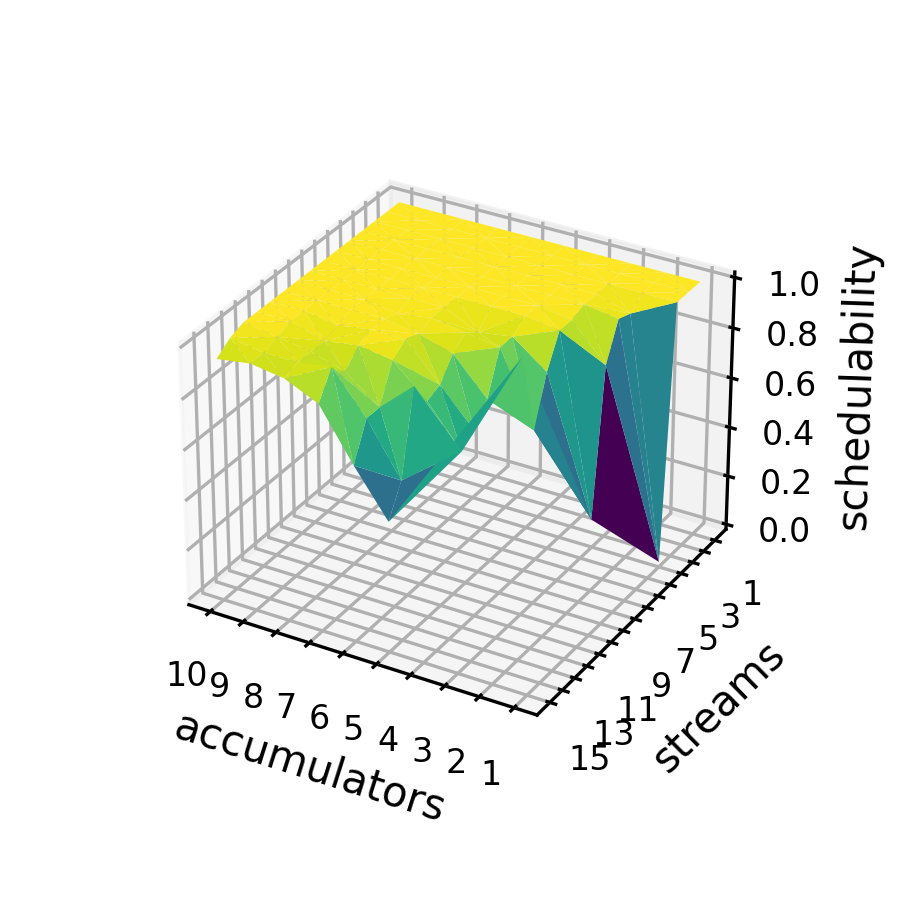}
	}%
	\subfloat[FF\label{fig:ami:success:FF}]{%
		\includegraphics[width=0.24\linewidth]{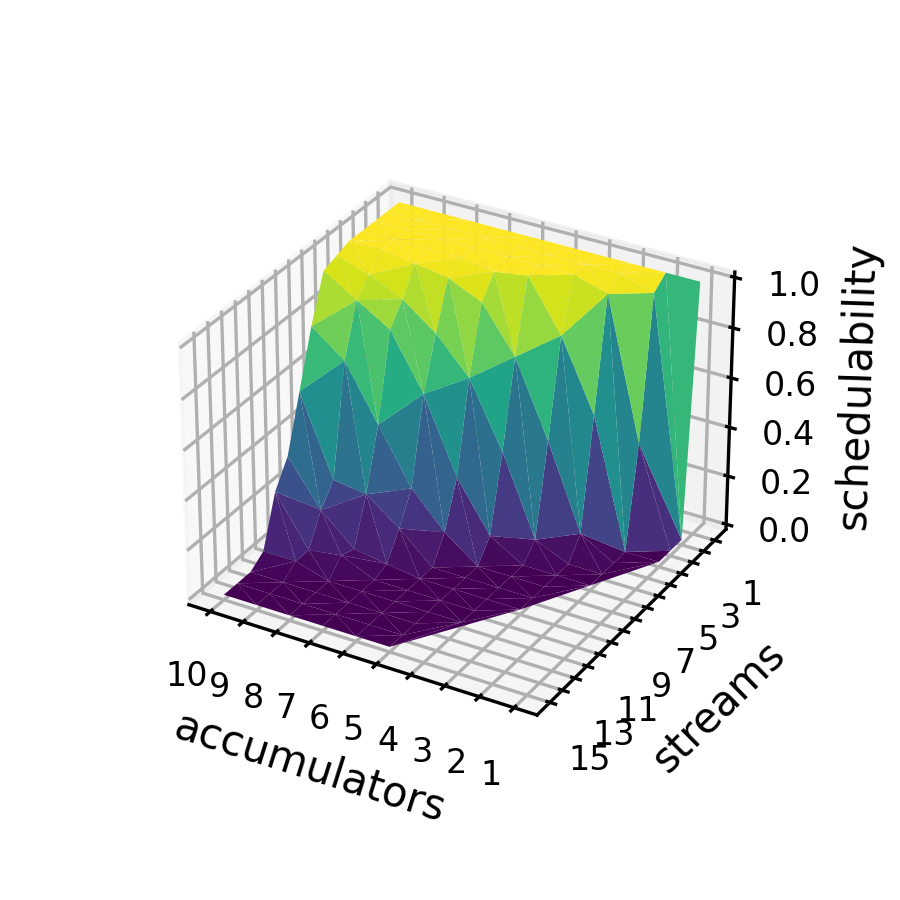}
	}%
	\subfloat[Hermes\label{fig:ami:success:Hermes}]{%
		\includegraphics[width=0.24\linewidth]{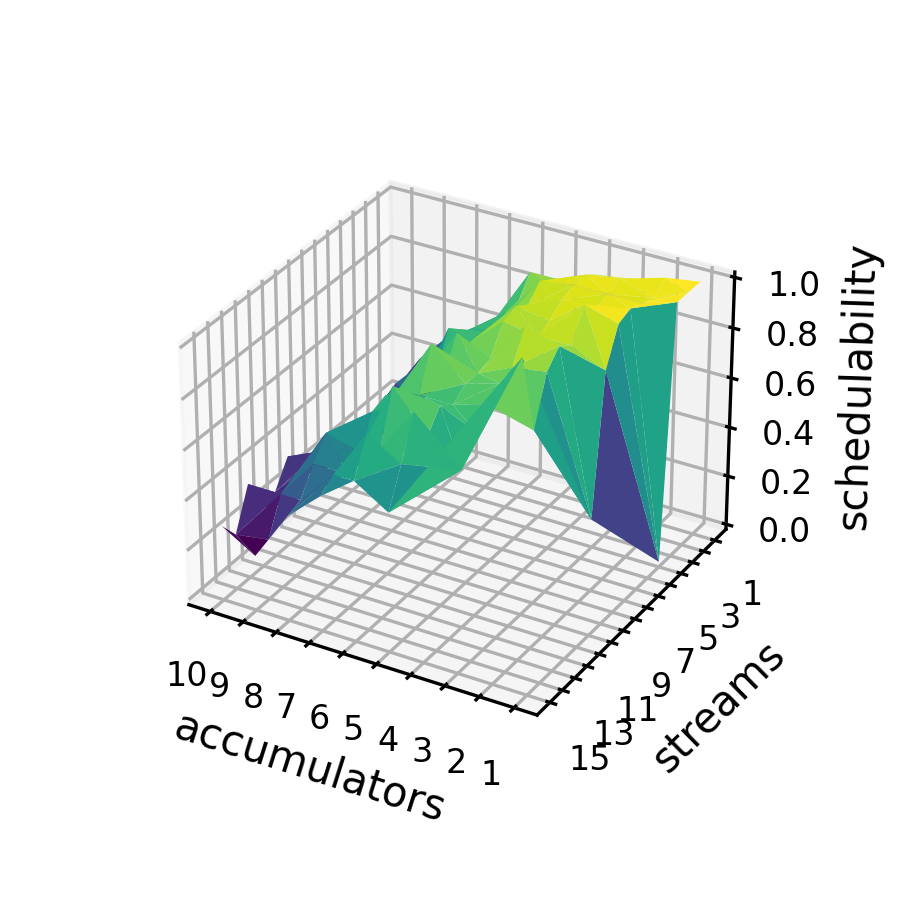}
	}
	%\vspace{-0.1cm}
	\caption{Schedulability in the AMI scenario with changing numbers of outgoing streams per concentrator and accumulators in the grid.}
	\label{fig:ami:success}
\end{figure*}

So far, we systematically evaluated the performance of H2S and CELF for a range of basic topologies, which occur in real networks, often in combination. 
In this case study, we went one step further and evaluated them in the context of a smart grid application. 
We used a synthetic electric grid model\footnote{IEEE 300-Bus System, available at \url{https://electricgrids.engr.tamu.edu/electric-grid-test-cases/}} developed by the IEEE Test Systems Task Force for grid performance and stability studies \cite{Birchfield2017}. 
This model has been designed to be statistically and functionally similar to actual electric grids.

As proposed in \cite{Zhu2019}, we assumed that the communication lines of the associated communication network are established along the landlines of the electrical grid network.
Further, we assumed  that every bus has a bridge connecting the landline edges.
Every generator, load, and transformer in the grid was modeled as sender/receiver, which is connected to the adjacent bus' bridge.
Typically, every bridge connected one or two end devices, i.e., receivers or senders.
Based on the underlying grid model, this lead to a communication network with 371 end devices, 405 bridges, and 887 network links.

We determined the traffic load for this communication network based on the communication demands of an Advanced Metering Infrastructure (AMI) application \cite{Gungor2013,Yan2013}, which continuously monitors the network state, using a time-triggered communication pattern.
An AMI application incorporates so-called accumulators (Meter Data Management Systems) and concentrators, where the latter simply bundle smart meter connections, while accumulators collect and process smart meter data from neighboring concentrators \cite{Lopez2015}.
In our evaluations, we varied the number of aggregators, which are randomly mapped to the end devices connected to the network.
Due to the lack of publicly available AMI data sets, we generated the streams based on load patterns reported in the literature:
The bidirectional communication between concentrators and accumulators were implemented as two time-triggered streams with an average throughput of \unit{500}{\kilo\bit\per\second} common for backhaul connections \cite{Gungor2013}.
The periods ranged from \unit{4-20}{\milli\second} covering the AMI latencies reported in \cite{Lopez2015,Yan2013}.
To account for synchronization and information exchange between the accumulators, we added a pair of time-triggered streams between any two accumulators with identical load characteristics.

We evaluated the scheduling strategies on the described network with different traffic pattern parameters. 
We varied the number of accumulators from 1 to 10 and had 1 to 15 outgoing (and incoming) streams per concentrator.
This allows us to model a wide range of metering bandwidth requirements to account for an increasing number of accumulators and higher volumes of monitoring data of future AMI applications \cite{Kerai2022}. 
We show the schedulability of H2S, CELF, FF, and Hermes in Figure~\ref{fig:ami:success}.
Note that we evaluated the schedulability instead of the aggregated throughput here.
It might not be reasonable to drop streams in a smart grid and allows for a fair comparison against Hermes.

H2S and CELF both showed a high degree of schedulability.
There were only a few scenarios where not all streams could be scheduled.
Still, in most cases 90\% of the streams were admitted.
Overall, H2S completely solved at least 90\% of the samples in 89.3\% of the accumulator/stream combinations, whereby the solving time was never above \unit{0.13}{\second}.
CELF achieved a comparable schedulability (at least 90\% of the samples in 87.7\% of the combinations).
In contrast to H2S, CELF needed up to \unit{1.4}{\second}.
If the aggregated bandwidth of the streams forwarded over a link comes close to the link's overall bandwidth capacity, it may happen that not all time constraints for those streams can be satisfied.
This happened in most scenarios that were not scheduled by H2S and CELF.
In contrast, FF failed to schedule most of the scenarios (90\% in only 33.6\% of the combinations), showing the benefit of the H2S stream and route sorting heuristics.
Hermes also yielded poor results (at least 90\% in only 20.5\% of the combinations) and required up to several minutes to compute a solution.
When comparing FF and Hermes specifically, we see that FF had issues with multiple streams, while Hermes had more trouble with additional accumulators.
The poor performance of Hermes was probably due to the algorithm running into deadlocks on multiple occasions.
Strategies like GFH and CP do not scale sufficiently well to solve problems of this size.
EDF does not yield better schedulability results than H2S, while the solving time ranges from a few seconds to several minutes.

In conclusion, we have seen that H2S is capable to solve the presented energy grid use case within much less than a second, while CELF provides comparable results in about a second.
Further, H2S outperformed its competitors, which had either less scheduling success, or required significantly more solving time (without yielding better schedulability).
The evaluation on the smart grid showed that the high aggregated throughput/scheduling success and low solving time of H2S that we already observed in the systematic topology evaluation come to full display in a realistic use case.
Therefore, H2S is a promising contender for real-world implementations with challenging topologies and varying network characteristics.

% !TeX spellcheck = en_US
% !TeX root = ../h2s_main.tex

\section{Conclusion}
\label{sec:conclusion}

In this paper, we considered the problem of scheduling and routing time-triggered streams in large TTEthernet networks, aiming for a high aggregated network throughput. 
We proposed two computationally efficient heuristics: \emph{Hierarchical Heuristic Scheduling} (H2S) and an adaptation \emph{Cost-efficient Lazy-forwarding Scheduling} (CELF).
Both are capable of scheduling large stream sets, yielding a high aggregated throughput on all evaluated networks while maintaining very fast solving times.
To be more specific, H2S scheduled over \unit{490}{\giga\bit\per\second} (more than 45,000 streams) on Erd\"os-R\'enyi random networks with 1,000 bridge nodes in less than a second, while the CELF scheduler required less than \unit{4}{\second}.
H2S and CELF clearly outperformed the other heuristics in terms of throughput, runtime, or both.
When compared to the CP-based solution, they yielded much better runtime and scalability.
For reasonably restricted time limits (in our case \unit{2}{\hour}), our heuristics even outperform CP solutions with respect to aggregated throughput.

% future work
In future work, we plan to extend the offensive versions of H2S and CELF Scheduling with more sophisticated defragmentation routines that avoid degradation of schedules over time.
In addition, the scheduling can be extended to multicast communications.
Therefore, the routing needs to be adapted to build candidate multicast trees instead of candidate routes, and the place method requires some changes.

\bibliographystyle{ieeetran}
\bibliography{h2s_main} 

% !TeX spellcheck = en_US
% !TeX root = ../h2s_main.tex

% The ieee main already has the \appendix call
\appendix
\section{Computing Candidate Routes}
\label{sec:routing}
\label{appendix:routing}

Routing can have a significant impact on the solution space of scheduling problems \cite{Nayak2018,Schweissguth2017}.
However, a completely free routing combined with scheduling increases the computational effort.
Therefore, a common solution is to use a set of routing options, so-called candidate routes \cite{Schweissguth2017}.
Using a k-shortest path algorithm like Yen's algorithm \cite{Yen1971} is a typical solution \cite{Falk2020,Falk2022,Pop2016}, with two main drawbacks:
First, the resulting routes can be very similar and thus, might suffer from the same congested links while there are still other short routes over different network links.
Second, computing k-shortest routes is a computationally expensive task, since the algorithm specifies that no other possible route is shorter than the returned k routes.
In contrast, for our use case, it is sufficient to determine k routes that are sort, but not necessarily the shortest ones.
Therefore, we used an adapted version of Dijkstra's shortest path routing that discourages (but does not ban) overlaps in the candidate routes.

The basic idea of our routing strategy is to compute new candidate routes iteratively with Dijkstra's shortest path algorithm.
For every computed route, the edge costs (initially 1) of the used links is increased for the following iterations.
Duplicate routes are omitted and when too many, i.e., 10, duplicates are found, the algorithms stops and yields less than k routes.
Thus, the runtime complexity results from the complexity of Dijkstra's algorithm multiplied by a factor dependent on the number of desired candidate routes.

Preliminary evaluations showed no clear difference in the resulting scheduling quality compared to yen's algorithm while having significantly shorter runtimes.
However, the exact impact of the candidate routing is not the focus of this paper and further evaluations are up to future work.

\section{Earliest Deadline First}
\label{appendix:edf}

Earliest Deadline First (EDF) is a well known scheduling method in single core CPU scheduling \cite{Kopetz2022}.
We use EDF as a benchmark, since it can be computed fast in reasonable sized scenarios and yields good results.
We simulate EDF to test its validity.
Thereby, we use the shortest route for all streams.
Otherwise, the runtime for EDF would not be feasible anymore due to the large configuration space.
First, we run a simulation with all streams.
If the first simulation fails, we start with smaller subset of the requested streams and then incrementally add streams in a FIFO order.
Therefore, we approximate the required network resources from the streams conservatively, by increasing their demand by 20%.
Thus, the resulting subset of streams should be definitely schedulable.

For every stream after the initial set, a new simulation is executed.
Note that every simulation will probably rearrange previous scheduling decisions and EDF is therefore an offensive approach. 
If the simulation fails, the newly added stream is rejected.
A single simulation can be performed very fast. 
However, as soon as the whole stream set becomes unschedulable, finding a large schedulable subset can be computationally expensive, because many streams will be added one by one, and thus need a lot of simulations.
It is important to note that we search for a large schedulable subset greedily and therefore cannot guarantee that we have the global maximum.

\section{Constraint Programming}
\label{appendix:cp}

To gain a ground truth for very small evaluation scenarios, we employ a Constraint Programming solution using IBM's CPLEX CP solver.
In the following, we describe the constraints, variables and optimization criteria used.

We have a one-dimensional array $S$ that a binary variable for every stream $s$ and stores a 1 ($S[s]=1$) whenever $s$ is admitted.
A second, two-dimensional, array R is used to determine which candidate route is used.
Hence, $R[s][r] = 1$ iff $s$ uses candidate route $r$.
A stream can use only a single candidate route which is ensured by the following constraint:

\begin{equation}
	\label{eq:cp:selected_path}
	\forall s \in S: \sum_r R[s][r] \leq 1
\end{equation}

The network link usage is stored in a map of arrays $L$.
Thereby, $L[(v,u)][s] = 1$ means the link between the network devices $v$ and $u$ is used by the selected candidate route of stream $s$ which is enforced via Constraint~\ref{eq:cp:link-usage}.
$P_r^s$ denotes the route $r$ of $s$ in the form of its consecutive network links.

\begin{equation}
	\label{eq:cp:link-usage}
	\forall (v,u) \in P_r^s: L[(v,u)][s] \geq R[s][r]
\end{equation}

The exact time slots when a stream occupies a link is described with an interval variable $I^{s,\text{seq.nr}_{(v,u)}}$ for every frame and every link it traverses.
The length of the interval is the frame's transmission delay and the start/end times are restricted by the release time/deadline. 
Further, the interval variables are set as optional, which allows the solver to reject a frame and therefore the whole stream.
We added a constraint to ensure that a stream is only considered admitted if all its frames are scheduled on all their hops:

\begin{equation}
	\label{eq:cp:admitted-flows}
	s \in S, \forall (v,u) \in P_r^s \forall \text{seq.nr} \in s: S[s] \leq \text{presence\_of}(I^{s,\text{seq.nr}}_{(v,u)})
\end{equation}

Thereby, $s$ denotes a stream of $S$ and might have multiple frames within a hyper period expressed via the $\text{seq.nr}$.
The links of the chosen candidate route $P_r^s$ are bespoken as the tuple $(v,u)$.
\emph{presence\_of} is a function of CPLEX which returns true (or 1) if the given variable is set.
Hence, the equation ensures that $s$ is only admitted ($S[s]$ is set to 1) iff all its frames are scheduled.

Next, we need to ensure that the frame transmissions within a single link $(v,u)$ do not interfere with each other.
CPLEX offers a \emph{no\_overlap} constraint which takes a group of Interval variables.
Thus, we grouped all frame transmissions $I^{s,\text{seq.nr}}_{(v,u)}$ by the network link and passed them to the no overlap constraint.
Finally, the frame ordering needs to be defined.
Otherwise, a frame might be sent via link $(w,u)$ before it completely traversed the previous network link $(v,u)$.
This is prevented by Constraint~\ref{eq:cp:ordering}:

\begin{equation}
	\label{eq:cp:ordering}
	\forall (v,w), (w,u) \in P^s: \text{end\_before\_start}(I^{s,\text{seq.nr}}_{(v,w)}, I^{s,\text{seq.nr}}_{(w,u)}, d)	
\end{equation}

\emph{end\_before\_start} denotes a CPLEX constraint that the first passed interval variable has to end at least $d$ time units before the second one starts.
Here, $d$ represents the propagation and processing delay.
Note that we can simply exchange the \emph{end\_before\_start} by \emph{end\_at\_start} to enforce zero queuing with the defined delay.

Finally, we need to define the optimization goal and include the streams traffic.
Therefore, we add an integer array $T$ with the same length as $S$ (one index per stream).
$T$ is filled using Constraint~\ref{eq:cp:traffic} so that every index stores 0 if the stream is rejected and the stream's throughput $thr(s)$ otherwise.
The optimization goal for the CP solver is to maximize the sum over $T$, which maximizes the aggregated traffic.
Note that we can easily derive a CP that optimizes the number of streams by maximizing the sum over $S$ instead.

\begin{equation}
	\label{eq:cp:traffic}
	T[s] = 
	\begin{cases}
		thr(s), & S[s] = 1\\
		0, & \text{otherwise}
	\end{cases}
\end{equation}

\end{document}